\documentclass[onecolumn,superscriptaddress,prd,aps,floats,floatfix,nofootinbib,amsmath,amssymb]{revtex4}
\usepackage{graphicx}
\usepackage{color}
\usepackage{amsmath,amssymb}
\usepackage{array}

\newcommand{\vev}[1]{\left\langle #1 \right\rangle}

\newcommand{\TeV}{\text{TeV}}
\newcommand{\GeV}{\text{GeV}}

\newcommand{\crit}{{\rm crit}}
\newcommand{\tr}{\mbox{tr}}
\newcommand{\gtwo}{I\kern-.1em I\,}
\newcommand{\be}{\begin{equation}}
\newcommand{\ee}{\end{equation}}
\newcommand{\beq}{\begin{eqnarray}}
\newcommand{\eeq}{\end{eqnarray}}
\newcommand{\bpm}{\begin{pmatrix}}
\newcommand{\epm}{\end{pmatrix}}
\newcommand{\cl}{\, \rm C.L.}

\begin{document}

\title{Higgs as a Top-Mode Pseudo}
\author{Hidenori S. Fukano}
\thanks{\tt fukano@kmi.nagoya-u.ac.jp}
      \affiliation{ Kobayashi-Maskawa Institute for the Origin of Particles and 
the Universe (KMI) \\ 
 Nagoya University, Nagoya 464-8602, Japan.}
\author{Masafumi Kurachi} \thanks{\tt kurachi@kmi.nagoya-u.ac.jp}
      \affiliation{ Kobayashi-Maskawa Institute for the Origin of Particles and 
the Universe (KMI) \\ 
 Nagoya University, Nagoya 464-8602, Japan.}
\author{Shinya Matsuzaki}\thanks{\tt synya@hken.phys.nagoya-u.ac.jp}
      \affiliation{ Institute for Advanced Research, Nagoya University, Nagoya 464-8602, Japan.}
      \affiliation{ Department of Physics, Nagoya University, Nagoya 464-8602, Japan.}
\author{{Koichi Yamawaki}} \thanks{
      {\tt yamawaki@kmi.nagoya-u.ac.jp}}
      \affiliation{ Kobayashi-Maskawa Institute for the Origin of Particles and 
the Universe (KMI) \\ 
 Nagoya University, Nagoya 464-8602, Japan.}

\begin{abstract}
In the spirit of the top quark condensation, 
we propose a model which has a naturally light composite Higgs boson, ``tHiggs" ($h^0_t$), 
to be identified with the 126 GeV Higgs discovered at the LHC. 
The tHiggs, a bound state of the top quark and its flavor (vector-like) partner, 
emerges as a pseudo Nambu-Goldstone boson (NGB), ``Top-Mode Pseudo'', 
together with the exact NGBs to be absorbed into the $W$ and $Z$ bosons as well as 
another (heavier) Top-Mode Pseudo (CP-odd composite scalar, $A^0_t$). 
Those five composite (exact/pseudo) NGBs are dynamically produced simultaneously 
by a single supercritical four-fermion interaction having $U(3) \times U(1)$ symmetry 
which includes the electroweak symmetry, 
where the vacuum is aligned by small explicit breaking term so as to break the symmetry down to 
a subgroup, $U(2) \times U(1)'$, 
in a way not to retain the electroweak symmetry, 
in sharp contrast to the little Higgs models. 
The explicit breaking term for the vacuum alignment gives rise to a mass of the tHiggs, 
which is  protected by the symmetry 
and hence naturally controlled against radiative corrections. 
Realistic top quark mass is easily realized similarly to the top-seesaw mechanism 
by introducing an extra (subcritical) four-fermion coupling which explicitly breaks 
the residual $U(2)' \times U(1)'$ symmetry with $U(2)'$ being an extra symmetry beside 
the above $U(3)_L \times U(1)$. 
We present a phenomenological Lagrangian of the Top-Mode Pseudos 
along with the standard model particles, 
which will be useful for the study of the collider phenomenology. 
The coupling property of the tHiggs is shown to be consistent 
with the currently available data reported from the LHC. 
Several phenomenological consequences and constraints from experiments are also addressed. 
\end{abstract}

\maketitle

\section{Introduction}

The ATLAS~\cite{Aad:2012tfa} and CMS collaborations~\cite{Chatrchyan:2012ufa} have discovered 
a new scalar particle at around 126 GeV having the properties 
compatible with the Higgs boson in the Standard Model (SM). 
However, the origin of mass is still a mysterious, 
since we do not yet understand detailed features of the 126 GeV Higgs, 
in particular the dynamical origin of the mass of the 126 GeV Higgs itself 
which is just a free parameter in the SM. 

A straightforward way to understand the dynamical origin of the Higgs boson 
in the explicit underlying theory beyond the SM is the walking technicolor 
having approximate scale invariance and large anomalous dimension $\gamma_m \simeq 1$, 
which predicts a technidilaton as a composite pseudo Nambu-Goldstone boson (NGB) 
of the approximate scale invariance~\cite{Yamawaki:1985zg,Bando:1986bg}. 
It in fact was shown to be consistent with the current LHC data 
for the 126 GeV Higgs~\cite{Matsuzaki:2012mk, Matsuzaki:2012xx}, 
and a recent lattice study~\cite{Aoki:2013qxa} showed 
indication of such a very light flavor-singlet scalar meson 
as a candidate for the technidilaton in the walking technicolor. 
In order to accommodate a large top quark mass in the walking technicolor, however, 
we would need even larger anomalous dimension $\gamma_m >1$ for the techni-condensate 
relevant to the top mass, possibly realized  by the additional strong four-fermion interaction 
(strong extended technicolor)~\cite{Miransky:1988gk,Matumoto:1989hf,Appelquist:1988as}. 
An alternative composite Higgs model based on such a strong four-fermion coupling is  
the top quark condensate model~\cite{Miransky:1988xi,Miransky:1989ds,Nambu:1989jt,Marciano:1989xd,Marciano:1989mj,Bardeen:1989ds} with $\gamma_m \simeq 2$. 
Here we propose a variant of the top quark condensate model based on the strong four-fermion couplings,
which yields a naturally light Higgs boson to be identified with  the 126 GeV Higgs boson at LHC.

Actually, among masses of the SM fermions, the top quark mass ($m_t \simeq 173 \,\GeV$) is 
the only one roughly of the order of the electroweak symmetry breaking (EWSB) scale 
($v_{_{\rm EW}} \simeq 246 \,\GeV$).  
Furthermore, the mass of the LHC Higgs boson ($m_h \simeq 126 \,\GeV$) 
is also roughly of the order of the EWSB scale. 
This coincidence may imply that the top quark plays a crucial role 
for both the EWSB and the generation of the mass of the Higgs boson. 
In fact, before the top quark was discovered with the mass being this large, 
the top quark condensation (Top-Mode Standard Model; TMSM) was proposed%
~\cite{Miransky:1988xi,Miransky:1989ds} 
to predict such a close relation among the top quark mass, 
the EWSB scale and the Higgs mass, 
based on the phase structure of the gauged Nambu-Jona-Lasinio (NJL) model%
~\cite{Kondo:1988qd,Appelquist:1988fm}. 
The four-fermion interactions in the TMSM are written 
in the SM-gauge-invariant form~\cite{Miransky:1988xi,Miransky:1989ds}:
\beq
{\cal L}^{4f}_{\rm TMSM}
=
G_t (\bar{q}_L^{i} t_R)({\bar t}_R {q_L}_i) + G_b (\bar{q}_L^{i} b_R)({\bar b}_R {q_L}_i ) 
+ 
G_{tb}(\bar{q}^i_L q^k_R)(i\tau^2)^{ij}(i\tau^2)^{kl}(\bar{q}^j_L q^l_R) 
+ \text{h.c.},
\label{TMSM-4fermi}
\eeq
where $q_{L(R)}=(t,b)^T_{L(R)}$ and $\tau^2$ is the second component of the Pauli matrices. 
It is straightforward to extend this to include 
all the three generations of the SM fermions~\cite{Miransky:1988xi,Miransky:1989ds}. 
The four-fermion interactions are arranged to trigger the top quark condensate, $\vev{\bar{t}t} \neq 0$, 
without other condensates such as the bottom condensate, 
$\vev{\bar{b}b}, \vev{\bar{t}b}, \cdots = 0$, in such a way that
\beq
 G_t> G_\crit >G_b, \cdots
 \,,
\eeq
where the critical coupling $G_\crit$ is given as $G_\crit = 4\pi^2/(N_c \Lambda^2)$, 
with $N_c$ being the number of QCD color and $\Lambda$ the cutoff scale of the theory, 
up to small corrections from the SM gauge interactions 
as implied by the phase structure of the gauged NJL model~\cite{Kondo:1988qd,Appelquist:1988fm}. 
The solution of the gap equation indicates that the top quark mass can be much smaller than 
the cutoff scale $\Lambda$, $m_t \ll \Lambda$, 
by tuning the four-fermion coupling close to the critical coupling, $0 < G_t/G_\crit -1 \ll 1$. 
The TMSM produces three NGBs which are absorbed into the $W$ and $Z$ bosons 
when the electroweak gauge interactions are switched on, 
and predicts the top quark mass to be on the order of the EWSB scale, 
$v_{_{\rm EW}} \simeq 246 \,\GeV$, 
through the Pagels-Stokar formula~\cite{Pagels:1979hd} 
for the decay constant $F_\pi (=v_{_{\rm EW}})$ of the NGBs, 
which are evaluated with the solution of the gap equation of the gauged NJL model. %

However, the original TMSM  has a few problems: 
i) Even if we assume the cutoff scale, $\Lambda$, is the Planck scale, 
the top quark mass is predicted to be $m_t = 220-250 \,\GeV$%
~\cite{Miransky:1988xi,Miransky:1989ds,Bardeen:1989ds}, 
which is somewhat larger than the experimental value 
$m^{\rm exp}_t = 173 \,\GeV$~\cite{Beringer:1900zz}. 
If we assume $\Lambda$ to be a few TeV to avoid excessive fine-tuning 
to reproduce the EWSB scale,
we would face a disastrous situation where the top quark mass is too large: 
$m_t \sim 600 \,\GeV$ (top mass problem); 
ii) the TMSM predicts a Higgs boson as a $t \bar{t}$ bound state (the ``top-Higgs boson'', $H_t$) 
with mass in a range of $m_t < m_{_{H_t}} < 2 m_t $. 
Such a top-Higgs boson cannot be identified as the Higgs boson with the mass $\simeq 126 \,\GeV$ 
which was discovered at the LHC~\cite{Aad:2012tfa,Chatrchyan:2012ufa} (Higgs mass problem). 

The top-seesaw model~\cite{Dobrescu:1997nm,Chivukula:1998wd,He:2001fz} 
can solve the top mass problem. 
In the top-seesaw model, 
a new (vector-like) $SU(2)_L$-singlet quark (seesaw partner of the top quark) is introduced 
to mix with the $t_R$, 
which pulls the top quark mass down to the desired value $\simeq 173\,\GeV$, 
so that the top mass problem is resolved. 
However, it turns out that the top-Higgs boson is still heavy, $m_t < m_{_{H_t}} < 2 m_t$, 
and therefore the Higgs mass problem still remains in the top-seesaw model. 

The Higgs mass problem was recently resolved 
by the top-seesaw assisted technicolor model~\cite{Fukano:2012qx,Fukano:2012nx}, 
which is a hybrid version combining the top-seesaw model with technicolor. 
In this model a light top-Higgs boson with $m_{_{H_t}} < m_t $ was realized 
by sharing the top quark mass with the technicolor sector. 
It was shown, however, that the coupling properties of the top-Higgs boson are quite different from 
those of the Higgs boson in the SM, 
and the model has currently been disfavored by the LHC data, 
most notably by the results on the diphoton decay channel~\cite{Aad:2012tfa,Chatrchyan:2012ufa} 
and production cross section through the vector boson fusion process%
~\cite{ATLAS-CONF-2013-034,CMS-PAS-HIG-13-005}. 

In this paper, in the spirit of the top quark condensation, 
we propose a model which solves the Higgs mass problem in a natural way, 
while keeping the solution of the top mass problem by the top-seesaw mechanism. 
The light composite Higgs, what we call ``tHiggs'', emerges as one of the composite pseudo NGBs, 
dubbed ``Top-Mode Pseudos", associated with the spontaneous breaking 
of an approximate global symmetry, 
which is triggered by strong (supercritical) four-fermion interactions. 

The model is constructed from the third generation quarks in the SM $q=(t,b)$ 
and a (vector-like) $\chi$-quark which is a flavor partner of the top quark, 
and has the same SM charges as those of the right-handed top quark. 
The four-fermion interaction term takes the form,
\beq
{\cal L}^{4f}  
= G (\bar{\psi}^i_L \chi_R)(\bar{\chi}_R \psi^i_L) 
\,,\label{tmP-4f}
\eeq
where $\psi^i_L \equiv (t_L , b_L, \chi_L)^{T\,i}\,,( i = 1,2,3)$. 
The four-fermion interaction in Eq.(\ref{tmP-4f}) possesses a global symmetry 
$U(3)_L \times U(1)_{\chi_R}$. 
For the supercritical setting, $G> G_\crit$, the symmetry is spontaneously broken down to 
$U(2)_L \times U(1)_V$ by the quark condensates $\vev{\bar{\chi}_R t_L} \neq 0$ 
and $\vev{\bar{\chi}_R \chi_L} \neq 0$, 
which are realized in the vacuum aligned with the additional explicit breaking terms mentioned below, 
while $\vev{\bar{\chi}_R b_L}$ is gauged away to $\vev{\bar{\chi}_R b_L} = 0$ 
by the electroweak gauge symmetry when it is switched on. 
Note that the electroweak gauge symmetry is spontaneously broken 
when $U(3)_L \times U(1)_{\chi_R} \to U(2)_L \times U(1)_V$, 
in sharp contrast to the little Higgs models. 
It should also be noted that the Lagrangian has $U(2)_{q_R}$ symmetry 
(see below Eq.(\ref{massless-top})) of $(t_R , b_R)$ not broken by the condensate, 
which will not be explicitly mentioned unless becomes relevant. 

Associated with this symmetry breaking, five NGBs emerge as bound states of the quarks. 
Besides those, a composite heavy Higgs boson corresponding to 
the $\sigma$ mode of the usual NJL model is also formed. 
Three of these NGBs will be eaten by the electroweak gauge bosons 
when the subgroup of the symmetry is gauged by the electroweak symmetry, 
while two of them remain as physical states. 
Those two NGBs, Top-Mode Pseudos, acquire their masses due to additional terms 
which explicitly break the $U(3)_L \times U(1)_{\chi_R}$ symmetry 
in such a way that the vacuum aligns to break the electroweak symmetry 
by $\vev{\bar{\chi}_R t_L} \neq 0$. 
One of them is a CP-even scalar (tHiggs, $h^0_t$), 
which is identified as the 126 GeV Higgs boson discovered at the LHC, 
while the other is a heavy CP-odd scalar ($A^0_t$), 
which is similar to the CP-odd Higgs in the two Higgs doublet models 
(there is an essential difference from the two-doublet Higgs model, though). 
We find a notable relation between masses of those Top-Mode Pseudos: 
\beq 
m_{h^0_t} = m_{A^0_t} \sin \theta
\,,\quad 
\left(\tan \theta  =\frac{\vev{\bar{\chi}_R t_L}}{ \vev{\bar{\chi}_R \chi_L}}\right)
\,,\label{mass-formula}
\eeq
where the angle $\theta$ is related to the presence of the condensate, 
$\vev{\bar{\chi}_R q_L} \neq 0$, 
which causes the electroweak symmetry breaking. 

It will be shown that the tHiggs couplings to the SM particles coincide with those of the SM Higgs boson 
in the limit $\sin \theta \to 0$ (where $v_{_{\rm EW}}  \simeq 246 \,\GeV$ is kept fixed). 
Even if the tHiggs coupling coincides with that of the SM Higgs, 
the virtue of our model is that the tHiggs $h^0_t$ is a bound state of the top quark and $\chi$-quark, 
and is natural in the sense that its mass is protected by the symmetry, 
in sharp contrast to the SM Higgs. 
One notable feature of our model is the prediction of the heavy CP-odd Higgs 
(without additional charged heavy Higgs in contrast to the two-doublet Higgs models), 
which will be tested in future collider experiments.  

This paper is organized as follows: 
In Sec.~\ref{Model}, we start with a simplified model based on four-fermion  dynamics 
which has the (exact) global symmetry $U(3)_L \times U(1)_{\chi_R}$ to show that 
five NGBs emerge due to the spontaneous breaking of the global symmetry 
by the quark condensate generated by the supercritical four-fermion  dynamics. 
Additional explicit breaking terms are then introduced to give masse to 
the two of the five NGBs (Top-Mode Pseudos; $h^0_t$ and $A^0_t$) and also to the top quark. 
We estimate the mass of the Top-Mode Pseudos based on the current algebra to find 
the mass formula Eq.(\ref{mass-formula}). 
The interaction property of the tHiggs ($h^0_t$) 
and the stability of the mass against the radiative corrections 
are addressed in comparison with the SM Higgs boson case. 
The extension of the model to incorporate masses of light fermions are also discussed 
in Sec.~\ref{Model}. 
Several phenomenological constraints are given in Sec.~\ref{phenomenology}. 
Sec.~\ref{summary} is devoted to the summary of this paper 
including some discussions.  
In appendix.~\ref{app-BHL}, we provide a straightforward derivation of the Top-Mode Pseudo masses 
by directly solving the bound state problem in the four-fermion dynamics 
based on the auxiliary field method. 
Appendix.~\ref{app-toploop-TMP} is devoted to  
some details of computations for one-loop corrections to 
the Top-Mode Pseudos arising from the top and its flavor partner, $t'$-quark loops.  


\section{Model}
\label{Model}
In this section we propose a model based on four-fermion dynamics constructed from 
the top and bottom quarks $q=(t,b)$ with the flavor partner of top quark $(\chi)$. 
The model possesses an approximate global symmetry which is spontaneously broken 
by the quark condensates $\vev{\bar{\chi}_R q_L} \neq 0$ and $\vev{\bar{\chi}_R \chi_L} \neq 0$ 
generated by the four-fermion dynamics.  
Five NGBs emerge as bound states of the quarks associated with the spontaneous breaking 
of the global symmetry, in addition to a composite  heavy Higgs boson 
which corresponds to the $\sigma$ mode in the usual NJL model. 
Two of the NGBs (what we call Top-Mode Pseudos ($h^0_t$ and $A^0_t$)) obtain their masses 
due to the introduction of additional four-fermion interactions which explicitly break the global symmetry, 
while other three remain massless to be eaten by the $W$ and $Z$ bosons 
once the electroweak charges are turned on. 
The mass of $h^0_t$ turns out to be protected by the global symmetry 
and the coupling property is shown to be consistent with the currently reported Higgs boson 
with mass around $\simeq 126\,\GeV$. We will call  $h^0_t$ tHiggs. 

The present model, which is based on the strong four-fermion interactions with vector-like $\chi$-quark,
can actually be viewed as a version of so-called top-seesaw model%
~\cite{Dobrescu:1997nm,Chivukula:1998wd}. 
The crucial difference between existing top-seesaw models and the present model is that 
the present model makes it clear that the 126 GeV Higgs exists as a pseudo NGB associated with 
the global symmetry breaking caused by four-fermion interactions. 
For the purpose of making this point clearer, in subsection~\ref{Model-DSB}, 
we introduce a simplified model in which all the explicit breaking terms are turned off. 
In that simplified model, five NGBs which exist in the model are all massless. 
Then, in subsection~\ref{TMPs}, we introduce explicit breaking terms into the Lagrangian 
to give masses to two of NGBs, Top Mode Pseudos ($h^0_t $, $A^0_t$), 
which are identified as 126 GeV Higgs boson and its CP-odd partner. 
Subsections~\ref{fermion-mass} and ~\ref{TNH} are devoted to explaining fermion masses, 
Yukawa interactions as well as the nature of the tHiggs.

\subsection{Structure of symmetry breaking}
\label{Model-DSB}

Let us consider an NJL-like model constructed from the third generation quarks in the SM, $q=(t,b)$, 
and an $SU(2)_L$ singlet quark $(\chi)$. 
The left-handed quarks $q_L$ and $\chi_L$ form a flavor triplet 
$\psi^i_L \equiv (t_L , b_L, \chi_L)^{T\,i}\, ( i = 1,2,3)$ under the flavor $U(3)_{\psi_L}$ group, 
while the right-handed top and bottom quarks $q^i_R \equiv (t_R , b_R)^{T\,i}\,(i=1,2)$ 
and $\chi_R$ are a doublet and singlet under the $U(2)_{q_R}$ group, respectively. 
Turning off the SM gauge interactions momentarily, we thus write the Lagrangian 
having the global $U(3)_{\psi_L} \times U(2)_{q_R} \times U(1)_{\chi_R}$ symmetry: 
\beq
{\cal L}_{\rm kin.} +
{\cal L}^{4f} 
&=& 
\bar{\psi}_L i \gamma^\mu \partial_\mu \psi_L
+
\bar{q}_R i \gamma^\mu \partial_\mu q_R
+
\bar{\chi}_R i \gamma^\mu \partial_\mu \chi_R
+
G (\bar{\psi}^i_L \chi_R)(\bar{\chi}_R \psi^i_L) 
\,,\label{start-Lag}
\eeq
where $G$ denotes the four-fermion coupling strength. 

We can derive the gap equations for fermion dynamical masses $m_{t\chi}$ and $m_{\chi\chi}$ 
through the mean field relations $m_{t\chi} = - G \vev{\bar{\chi}_R t_L}$ 
and $m_{\chi\chi} = - G \vev{\bar{\chi}_R \chi_L}$ in the large $N_c$ limit:\footnote{
We have put $m_{b\chi} =0$, by gauging  it away by the electroweak gauge symmetry. 
Otherwise $(m_{t\chi}^2+m_{\chi\chi}^2)$ in Eqs.(\ref{gapeq-tchi-0}) and (\ref{gapeq-chichi-0}) 
should read $(m_{t\chi}^2+m_{b\chi}^2+m_{\chi\chi}^2)$ 
because of $U(3)_{\psi_L}$ symmetry, 
with $m_{b\chi}$ also satisfying the same type of gap equation.
}
\beq
m_{t\chi} 
&=&
m_{t\chi} \frac{N_c G }{8\pi^2} 
\left[
\Lambda^2 -
\left( m^2_{t\chi} + m^2_{\chi \chi} \right) \ln \frac{\Lambda^2}{m^2_{t\chi} + m^2_{\chi\chi}}
\right]
\,,\label{gapeq-tchi-0}
\\
m_{\chi\chi} 
&=&
m_{\chi\chi} \frac{N_c G }{8\pi^2} 
\left[
\Lambda^2 -
\left( m^2_{t\chi} + m^2_{\chi \chi} \right) \ln \frac{\Lambda^2}{m^2_{t\chi} + m^2_{\chi\chi}}
\right]
\,,\label{gapeq-chichi-0}
\eeq
where $\Lambda$ stands for the cutoff of the model which is to be of 
$\Lambda \gg {\cal O}(1)\,\TeV$. 
There exist nontrivial solutions $m_{t\chi} \neq 0$ and $m_{\chi\chi} \neq 0$ 
when the following criticality condition is satisfied:  
\beq
G > G_\crit = \frac{8\pi^2}{N_c \Lambda^2}\, , 
\label{crticality-condition}
\eeq 
under which we have the nonzero dynamical masses as well as the nonzero condensates, 
\beq
\vev{\bar{\chi}_R q_L} \neq 0 
\,, \qquad
\vev{\bar{\chi}_R \chi_L} \neq 0 
\,. 
\label{condensation-psiBasis}
\eeq 
Note, however, that the two gap equations, Eqs.(\ref{gapeq-tchi-0}) and (\ref{gapeq-chichi-0}), 
cannot determine the ratio of two condensates: 
Those two gap equations with the criticality condition in Eq.(\ref{crticality-condition}) 
just lead to the nontrivial solution 
for the squared-sum of two masses, $(m^2_{t \chi} + m^2_{\chi\chi}) \neq 0$, 
so that the vacuum with $m_{t \chi} \neq 0$ is degenerate with that with $m_{t \chi} = 0$. 
In order to lift the degeneracy for breaking the electroweak symmetry by $m_{t \chi} \neq 0$, 
we shall later introduce explicit breaking to align the vacuum,  
which also gives rise to the mass of two NGBs (Top-Mode Pseudos) out of five NGBs, 
with the rest three being exact NGBs to be absorbed into the $W$ and $Z$ bosons.  
Also note that the condensate $m_{b \chi} \neq 0$ can be gauged away 
when the model is gauged by the electroweak symmetry. 

In order to make the structure of the symmetry breaking clearer, 
we may change the flavor basis of fermions by introducing an orthogonal rotation matrix $R$:  
\beq
\tilde{\psi}_L
=
\bpm
\tilde{t}_L \\[1ex] \tilde{b}_L \\[1ex] \tilde{\chi}_L
\epm
\equiv
R \cdot \psi_L
\quad , \quad
R
=
\bpm
\cos \theta & 0 & - \sin \theta \\[1ex]
0 & 1 & 0 \\[1ex]
\sin \theta & 0 & \cos \theta 
\epm
\quad ,\quad
\tan \theta \equiv \frac{m_{t\chi}}{m_{\chi\chi}} 
=
\frac{\vev{\bar{\chi}_R t_L}}{ \vev{\bar{\chi}_R \chi_L}}. 
\label{rotation-psiL}
\eeq
The two gap equations, Eqs.(\ref{gapeq-tchi-0}) and (\ref{gapeq-chichi-0}), 
are then reduced to a single gap equation, 
\beq
1 
&=&
\frac{N_c G }{8\pi^2} 
\left[
\Lambda^2 -
m^2_{\tilde{\chi}\chi}
\ln \frac{\Lambda^2}{m^2_{\tilde{\chi}\chi}}
\right]
\,,\label{reduced-gapeq-0}
\eeq
with
\beq
m^2_{\tilde{\chi}\chi} \equiv m^2_{t\chi} + m^2_{\chi\chi} \neq 0
\,. \label{dynamical-mass-psitilde}
\eeq
Accordingly, the associated two condensates in Eq.(\ref{condensation-psiBasis}) 
are reduced to a single nonzero condensate on the basis of $\tilde{\psi}_L$: 
\beq
\vev{\bar{\chi}_R \tilde{\chi}_L} \neq 0
\,.\label{condensation-tild}
\eeq
We thus see that, with the criticality condition in Eq.(\ref{crticality-condition}) satisfied, 
the four-fermion dynamics triggers the following global symmetry breaking pattern:  
\beq
U(3)_{\tilde{\psi}_L} \times U(1)_{\chi_R} 
\to 
U(2)_{\tilde{q}_L} \times U(1)_{V=\tilde{\chi}_L + \chi_R} 
\,.\label{TMP-G/H}
\eeq
The broken currents associated with this symmetry breaking are found to be 
\beq
J^{4,\mu}_{3L}
&=&
\bar{\tilde{\psi}}_L \gamma^\mu \lambda^4 \tilde{\psi}_L
\nonumber\\
&=&
\bar{\tilde{t}}_L \gamma^\mu \tilde{\chi}_L + \bar{\tilde{\chi}}_L \gamma^\mu \tilde{t}_L 
\nonumber\\
&=&
\left( \bar{t}_L \gamma^\mu t_L + \bar{\chi}_L \gamma^\mu \chi_L \right) \sin 2\theta
+
\left( \bar{t}_L \gamma^\mu \chi_L + \bar{\chi}_L \gamma^\mu t_L \right) \cos 2\theta
\,,\label{broken-current-3L4}
\\[1ex]
J^{5,\mu}_{3L}
&=&
\bar{\tilde{\psi}}_L \gamma^\mu \lambda^5\tilde{\psi}_L
\nonumber\\
&=&
 i \left[ 
-\bar{\tilde{t}}_L \gamma^\mu \tilde{\chi}_L 
+ 
\bar{\tilde{\chi}}_L \gamma^\mu \tilde{t}_L \right]
\nonumber\\
&=&
-i\left( \bar{t}_L \gamma^\mu \chi_L - \bar{\chi}_L \gamma^\mu t_L \right) 
\,,\label{broken-current-3L5}
\\[1ex]
J^{6,\mu}_{3L}
&=&
\bar{\tilde{\psi}}_L \gamma^\mu \lambda^6\tilde{\psi}_L
\nonumber\\
&=&
 \bar{\tilde{b}}_L \gamma^\mu \tilde{\chi}_L + \bar{\tilde{\chi}}_L \gamma^\mu \tilde{b}_L 
\nonumber\\
&=&
\left( \bar{b}_L \gamma^\mu t_L + \bar{t}_L \gamma^\mu b_L \right) \sin \theta
+
\left( \bar{b}_L \gamma^\mu \chi_L + \bar{\chi}_L \gamma^\mu b_L \right) \cos \theta
\,,\label{broken-current-3L6}
\\[1ex]
J^{7,\mu}_{3L}
&=&
\bar{\tilde{\psi}}_L \gamma^\mu \lambda^7 \tilde{\psi}_L
\nonumber\\
&=& 
i \left[ 
-\bar{\tilde{b}}_L \gamma^\mu \tilde{\chi}_L 
+ 
\bar{\tilde{\chi}}_L \gamma^\mu \tilde{b}_L \right]
\nonumber\\
&=&
-i \left( \bar{b}_L \gamma^\mu t_L - \bar{t}_L \gamma^\mu b_L \right) \sin \theta
-i \left( \bar{b}_L \gamma^\mu \chi_L - \bar{\chi}_L \gamma^\mu b_L \right) \cos \theta
\,,\label{broken-current-3L7}
\eeq
and 
\beq
J^\mu_A 
&\equiv& 
\frac{1}{4}
\left(
J^\mu_{1R} - \frac{1}{\sqrt{6}} J^{0,\mu}_{3L} + \frac{1}{\sqrt{3}} J^{8,\mu}_{3L}
\right)
\nonumber\\
&=&
\frac{1}{4}
\left(
\bar{\chi}_R \gamma^\mu \chi_R - \bar{\tilde{\chi}}_L \gamma^\mu \tilde{\chi}_L
\right)
\nonumber\\
&=&
\frac{1}{4} \left[
\bar{\chi}_R \gamma^\mu \chi_R 
- \bar{t}_L \gamma^\mu t_L \sin^2\theta
- \bar{\chi}_L \gamma^\mu \chi_L \cos^2\theta
- (\bar{t}_L \gamma^\mu \chi_L + \bar{\chi}_L \gamma^\mu t_L) \sin\theta\cos\theta
\right]
\,,\label{broken-current-A}
\eeq 
where the Gell-Mann matrices $\lambda^a\,(a=1, \cdots ,8)$ are normalized as 
${\rm tr}[\lambda^a \lambda^b]=2 \delta^{ab}$, and $\lambda^0 = \sqrt{2/3} \,\, 1_{3\times 3}$. 
The associated NGBs emerge with the decay constant $f$ as  
\beq 
\left \langle 0 \left |
J^a_\mu(x)
\right| \pi^b_t (p)\right \rangle  
=
- i f \delta^{ab} p_\mu e^{-i p\cdot x} 
\,,  
\qquad 
a, b = 4, 5,6,7, A  
\,. \label{def-decay-const}
\eeq
The decay constant $f$ is calculated through the Pagels-Stokar formula~\cite{Pagels:1979hd}: 
\beq
f^2 = \frac{N_c}{8\pi^2} m^2_{\tilde{\chi}\chi} \ln\frac{\Lambda^2}{m^2_{\tilde{\chi}\chi}}
\,.\label{model-PSformula}
\eeq 
The five NGBs $(\pi^a_t)$ can be expressed as composite fields (interpolating fields) 
made of the fermion bilinears on the basis of ($\tilde{\psi}_L$, $\chi_R$) or ($\psi_L, \chi_R$):   
\beq
\pi^4_t
&\sim&
\bar{\chi}_R \tilde{t}_L - \bar{\tilde{t}}_L \chi_R
\nonumber\\
&=&
\left( \bar{\chi}_R t_L - \bar{t}_L \chi_R \right) \cos\theta
-
\left( \bar{\chi}_R\chi_L - \bar{\chi}_L \chi_R\right) \sin \theta
\nonumber
\,,\\
\pi^5_t
&\sim&
-i \left( \bar{\chi}_R \tilde{t}_L + \bar{\tilde{t}}_L \chi_R \right)
\nonumber\\
&=&
-i \left( \bar{\chi}_R t_L + \bar{t}_L \chi_R \right) \cos\theta
+
i \left( \bar{\chi}_R\chi_L + \bar{\chi}_L \chi_R\right) \sin \theta
\nonumber
\,,\\
\pi^6_t + i \pi^7_t
&\sim&
\left(\bar{\chi}_R \tilde{b}_L - \bar{\tilde{b}}_L \chi_R\right) 
+ 
\left( \bar{\chi}_R \tilde{b}_L + \bar{\tilde{b}}_L \chi_R \right)
\nonumber\\
&=&
2 \bar{\chi}_R b_L 
\nonumber
\,,\\
\pi^6_t - i \pi^7_t
&\sim&
\left(\bar{\chi}_R \tilde{b}_L - \bar{\tilde{b}}_L \chi_R\right) 
- 
\left( \bar{\chi}_R \tilde{b}_L + \bar{\tilde{b}}_L \chi_R \right)
\nonumber\\
&=&
-2 \bar{b}_L \chi_R 
\nonumber
\,,\\
\pi^A_t
&\sim&
\bar{\chi}_R \tilde{\chi}_L - \bar{\tilde{\chi}}_L \chi_R
\nonumber\\
&=&
\left( \bar{\chi}_R t_L - \bar{t}_L \chi_R \right) \sin\theta
+
\left( \bar{\chi}_R\chi_L - \bar{\chi}_L \chi_R\right) \cos \theta
\nonumber
\,.
\eeq
Besides these composite NGBs, there exists a composite scalar ($H^0_t$) 
corresponding to the $\sigma$ mode in the usual NJL model,  
\beq
H^0_t 
&\sim& 
\bar{\chi}_R \tilde{\chi}_L  + \bar{\tilde{\chi}}_L \chi_R
\nonumber\\
&=& 
\left( \bar{\chi}_R t_L + \bar{t}_L \chi_R \right) \sin\theta
+
\left( \bar{\chi}_R\chi_L + \bar{\chi}_L \chi_R\right) \cos \theta
\,,\nonumber
\eeq
with the mass  
\beq
m^2_{H^0_t} = 4 m^2_{\tilde{\chi}\chi} = 4 (m^2_{t\chi} + m^2_{\chi\chi})
\,.\label{mass-Ht0}
\eeq 
The $H^0_t$ will be regarded as a heavy Higgs boson with the mass of ${\cal O}(1)$ TeV, 
not the light Higgs boson at around 126 GeV. 

With the electroweak gauge interactions turned on, 
the $W$ and $Z$ bosons turn out to couple to 
the broken currents $(J_{3L}^{6\,,\mu} \mp i J_{3L}^{\mu\,,7})$ 
and $(J_{3L}^{\mu\,,4} \cos\theta + J_{3L}^{\mu\,,A} \sin\theta)$, respectively. 
The corresponding would-be NGBs ($w^\pm_t, z^0_t$) eaten 
by the electroweak gauge bosons are then found to be 
\beq
z^0_t 
&\equiv& 
\pi^4_t \cos\theta + \pi^A_t \sin\theta
\nonumber\\
&\sim&
\bar{\chi}_R t_L - \bar{t}_L \chi_R
\,,\nonumber
\\
w^-_t &\equiv& 
\frac{1}{\sqrt{2}} (\pi^6_t + i \pi^7_t )
\nonumber\\
&\sim&
\sqrt{2} \bar{\chi}_R b_L
\,,\nonumber
\\
w^+_t &\equiv& 
\frac{1}{\sqrt{2}} (\pi^6_t - i \pi^7_t )
\nonumber\\
&\sim&
-\sqrt{2} \bar{b}_L \chi_R
\,. \nonumber
\eeq
On the other hand, the following two NGBs remain as physical states:  
\beq
h^0_t &\equiv& \pi^5_t 
\nonumber \\ 
&\sim&
-i \left( \bar{\chi}_R \tilde{t}_L + \bar{\tilde{t}}_L \chi_R \right)
\nonumber\\
&=&
-i \left( \bar{\chi}_R t_L + \bar{t}_L \chi_R \right) \cos\theta
+
i \left( \bar{\chi}_R\chi_L + \bar{\chi}_L \chi_R\right) \sin \theta 
\nonumber\,, \\ 
A^0_t 
&\equiv& 
- \pi^4_t \sin\theta + \pi^A_t \cos\theta
\nonumber \\ 
&\sim& 
\bar{\chi}_R \chi_L - \bar{\chi}_L \chi_R
\,.\nonumber
\eeq
The correspondence between the broken currents and NGBs 
along with the CP transformation property 
is summarized in Table.\ref{NGB-correspondence}. 
The two massless NGBs $(h^0_t, A^0_t)$ will become pseudo NGBs, called ``Top-Mode Pseudos", 
obtaining their masses once explicit breaking effects are introduced (see Sec.~\ref{TMPs}). 
We will identify the CP-even Top-Mode Pseudo, $h^0_t$, as the 126 GeV Higgs, called tHiggs. 

\begin{table}[h]
\begin{center}
\begin{tabular}{| c | c  | c |}
\hline
\hspace{15pt}  Broken current \hspace{15pt}  & \hspace{20pt} corresponding NGB \hspace{20pt} & \hspace{15pt}  CP-property \hspace{15pt}  
\\
\hline
\parbox[c][5ex][c]{0ex}{}
$J^{4,\mu}_{3L}$ 
& $\pi^4_t = z^0_t \cos\theta-A^0_t \sin\theta$  & odd 
\\
\hline 
\parbox[c][5ex][c]{0ex}{}
$J^{5,\mu}_{3L}$  
& $\pi^5_t = h^0_t$ & even 
\\
\hline
\parbox[c][5ex][c]{0ex}{}
$J^{6,\mu}_{3L} \pm i J^{7,\mu}_{3L}$ 
& $\pi^6_t \pm i \pi^7_t = \sqrt{2} w^\mp_t$& -- 
\\ 
\hline
\parbox[c][5ex][c]{0ex}{}
$J^\mu_A$
& $\pi^A_t = z^0_t \sin\theta + A^0_t\cos\theta$  &  odd 
\\ 
\hline
\end{tabular}
\caption{
The list of the broken currents and NGBs associated with the spontaneous symmetry breaking 
in Eq.(\ref{TMP-G/H}). 
$w^\pm_t$ and $z^0_t$ are eaten by  the $W^\pm$ and $Z$ bosons 
once the electroweak gauges are turned on, 
while the remaining two  $A^0_t$ and  $h^0_t$ become pseudo NGBs (Top-Mode Pseudos) 
by explicit breaking effects (see Eqs.(\ref{mass-At0}) and (\ref{mass-ht0})).  
} 
\label{NGB-correspondence}
\end{center}
\end{table}%

We may integrate out the heavy Higgs boson $H^0_t$ (with mass of ${\cal O}(1)$ TeV) 
to construct the low-energy effective theory governed 
by the five NGBs $(\pi^a_t)$ described by a nonlinear sigma model based on the coset space, 
\beq
\frac{\cal G}{\cal H}
=
\frac{U(3)_{\tilde{\psi}_L} \times U(1)_{\chi_R}}{U(2)_{\tilde{\psi}_L} \times U(1)_{V=\chi_L + \chi_R}}
\,.\label{TMP-coset}
\eeq
 For this purpose we introduce representatives ($\xi_{L,R}$) of the ${\cal G}/{\cal H}$ 
 which are parameterized by NGB fields as
\beq
\xi_L = 
\exp\left[ 
- \frac{i}{f} \left( 
\sum_{a=4,5,6,7} \pi^a_t \lambda^a  + \frac{\pi^A_t}{2\sqrt{2}} \lambda^A 
\right) 
\right]
\quad , \quad
\xi_R = 
\exp\left[ 
\frac{i}{f} 
\frac{\pi^A_t}{2\sqrt{2}} \lambda^A 
\right]
\,,\nonumber
\eeq
where
\beq
\lambda^A
= 
\bpm 0 & 0 & 0 \\ 0 & 0 & 0 \\ 0 & 0 & \sqrt{2} \epm
\,.\nonumber
\eeq
We further introduce the ``chiral" field $U$,
\beq
U = \xi_L^\dag \cdot \Sigma \cdot \xi_R 
\quad
\text{with \,\, $\Sigma = \frac{1}{\sqrt{2}} \lambda^A$}
\,.\label{def-NLsM-U}
\eeq
The transformation properties of $\xi_{L,R}$ and $U$ under ${\cal G}$ are given by
\beq
\xi_L \to  h(\pi_t,\tilde{g}) \cdot\xi_L \cdot g_{\tilde{3L}}^\dag
\quad , \quad
\xi_R \to  h(\pi_t,\tilde{g}) \cdot\xi_R \cdot g_{1R}^\dag
\quad , \quad
U \to g_{\tilde{3L}} \cdot U \cdot g^\dagger_{1R}
\,,\label{transform-NLsM-U}
\eeq
where $\tilde{g}=\{ g_{\tilde{3L}},g_{1R}\}\,,\,g_{\tilde{3L}} \in U(3)_{\tilde{\psi}_L} \,,\,g_{1R} \in U(1)_{\chi_R}$ 
and $h(\pi_t,\tilde{g}) \in {\cal H}$. 
Thus we find the ${\cal G}$-invariant Lagrangian written 
in terms of the NGBs to the lowest order of derivatives of ${\cal O}(p^2)$: 
\beq
{\cal L}_{\text{NL$\sigma$M}}
=
\frac{f^2}{2} \tr\left[  \partial_\mu U^\dagger  \partial^\mu U\right]
\,.
\label{NLsM-Lag-Op2-0}
\eeq%
When the electroweak symmetry turned on, the covariant derivative acting on $U$ is given by 
\beq
D_\mu U
\equiv
R \left[
\partial_\mu 
- i g \sum^3_{a=1} W^a_\mu 
\left(
\begin{array}{cc|c}
&&0\\
\mbox{\raisebox{1.5ex}{\smash{\large$\mspace{5mu}\tau^a/2$}}}&&0\\ \hline
\mspace{-20mu}0&\mspace{-20mu}0&0
\end{array}
\right)
+ i g' B_\mu \bpm 1/2 & 0 & 0 \\ 0 & 1/2 & 0 \\ 0 & 0 & 0 \epm 
\right]
R^T
\cdot
U 
\,,\label{covariant-derivative-U}
\eeq 
where $W_\mu$ and $B_\mu$ are the $SU(2)_L$ and $U(1)_Y$ gauge boson fields 
with the gauge couplings $g$ and $g'$. 
Then the Lagrangian Eq.(\ref{NLsM-Lag-Op2-0}) is changed to the covariant form:
\beq
{\cal L}_{\text{NL$\sigma$M}}
=
\frac{f^2}{2} \tr\left[  D_\mu  U^\dagger  D^\mu  U\right]
\,.\label{NLsM-Lag-Op2}
\eeq
From this one can read off the $W$ an $Z$ boson masses as  
\beq
m^2_W =  \frac{1}{4}g^2 f^2 \sin^2\theta
\quad , \quad 
m^2_Z = \frac{1}{4}(g^2 + g'^2) f^2 \sin^2\theta
\,,\nonumber
\eeq
which lead to 
\beq
v^2_{_{\rm EW}}
&=& f^2\sin^2\theta 
\nonumber\\
&=&
\frac{N_c}{8\pi^2} m^2_{t\chi} \ln\frac{\Lambda^2}{m_{t\chi}^2 + m_{\chi\chi}^2}
\nonumber\\
&\simeq &
(246 \,\GeV)^2 
\,,\label{def-vEW}
\eeq
where use has been made of Eq.(\ref{model-PSformula}). 
Thus imposing the EWSB scale $v_{_{\rm EW}}$ gives a nontrivial relation 
between $m_{t\chi}$ and $m_{\chi\chi}$. 

Note that switching on the electroweak gauge interaction explicitly breaks 
the $U(3)_{\tilde{\psi}_L} \times U(1)_{\chi_R}$ symmetry. 
Such explicit breaking effects would generate masses to the NGBs at the loop level, 
as a part of the $1/N_c$ sub-leading effect,
which are, however, negligibly small since the size of effects is suppressed 
by the small electroweak gauge coupling, 
as will be discussed later 
(see the discussion below Eq.(\ref{t-prime-integ-mh2})).

As we mentioned earlier the criticality $G>G_\crit$ implies 
$\vev{\bar{\chi}_R t_L}^2+ \vev{\bar{\chi}_R \chi_L}^2 \neq 0$, 
but not necessarily $\vev{\bar{\chi}_R t_L} \neq 0$ 
which is responsible for the electroweak symmetry breaking.  
The electroweak gauge interaction itself can contribute to lifting the degeneracy 
between the vacuum with $\vev{\bar{\chi}_R t_L} = 0$ and that with $\vev{\bar{\chi}_R t_L} \neq 0$ 
in principle by some extreme fine tuning of the critical coupling 
of the gauged NJL model~\cite{Kondo:1988qd,Appelquist:1988fm}. 

More natural way will be to introduce extra effective four-fermion interactions to explicitly break 
the $U(3)_{\tilde{\psi}_L} \times U(1)_{\chi_R}$ symmetry, 
which can align the vacuum to have $\vev{\bar{\chi}_R t_L} \ne 0$ 
and simultaneously give the mass of the tHiggs on the right amount. 
Such an explicit breaking may be induced from a strong $U(1)$ gauge interaction 
distinguishing $\vev{\bar{\chi}_R t_L}$ from $\vev{\bar{\chi}_R \chi_L}$. 
This we perform in the next subsection.

\subsection{Top-Mode Pseudos}
\label{TMPs}

Here we incorporate explicit breaking terms into the Lagrangian Eq.(\ref{start-Lag}) 
to give masses to the Top-Mode Pseudos $(h^0_t, A^0_t)$:   
\beq
{\cal L}_{\rm kin.} + {\cal L}^{4f} + {\cal L}^h 
\,,\label{with-explicit-breaking-00}
\eeq
where 
\beq
{\cal L}^h
=
-
\left[\Delta_{\chi \chi} \bar{\chi}_R \chi_L
+ \text{h.c.}
\right]
- G' \left( \bar{\chi}_L \chi_R \right) \left( \bar{\chi}_R \chi_L \right)
\,. \label{with-explicit-breaking-0}
\eeq
  
Similarly to Eqs.(\ref{gapeq-tchi-0}) and (\ref{gapeq-chichi-0}), 
we derive the gap equations 
for fermion dynamical masses $m_{t\chi}$ and $m_{\chi\chi}$:
\beq
m_{t\chi} 
&=&
m_{t\chi} \frac{N_c G }{8\pi^2} 
\left[
\Lambda^2 -
m^2_{\tilde{\chi}\chi} \ln \frac{\Lambda^2}{m^2_{\tilde{\chi}\chi}}
\right]
\,,\label{gapeq-tchi}
\\[1ex]
m_{\chi\chi} 
&=&
\Delta_{\chi\chi}
+
m_{\chi\chi} \frac{N_c (G-G') }{8\pi^2} 
\left[
\Lambda^2 -
m^2_{\tilde{\chi}\chi} \ln \frac{\Lambda^2}{m^2_{\tilde{\chi}\chi}}
\right]
\,, \label{gapeq-chichi}
\eeq
where $m^2_{\tilde{\chi}\chi} = m^2_{t\chi} + m^2_{\chi\chi}$. 
Note that the nonzero $\Delta_{\chi\chi}$ and $G'$ allow to determine 
the ratio of two dynamical masses $m_{t\chi}$ and $m_{\chi\chi}$, 
i.e. $\tan \theta =m_{t\chi}/m_{\chi\chi}$, 
in contrast to the previous gap equations, Eqs.(\ref{gapeq-tchi-0}) and (\ref{gapeq-chichi-0}), 
which only determine the squared-sum of two, $m^2_{t\chi} + m^2_{\chi\chi}$.

Furthermore, it turns out that these $\Delta_{\chi\chi}$ and $G'$ terms do not affect 
the criticality of the four-fermion dynamics at all: 
Assuming $m_{t\chi} \neq 0$ and $m_{\chi\chi}\neq 0$, 
we find that the following relation is required so as to keep the self consistency in the gap equations: 
\beq
\Delta_{\chi\chi}
=
\frac{G'}{G} m_{\chi \chi}
=
-G'\vev{\bar{\chi}_R \chi_L}
\,.
\label{relation-Gapeq}
\eeq
By taking this into account, 
the two gap equations, Eqs.(\ref{gapeq-tchi}) and (\ref{gapeq-chichi}), 
are reduced to a single one, 
\beq
1 
&=&
\frac{N_c G }{8\pi^2} 
\left[
\Lambda^2 -
m^2_{\tilde{\chi}\chi} \ln \frac{\Lambda^2}{m_{\tilde{\chi}\chi}^2}
\right]
\,.\label{reduced-gapeq}
\eeq
Thus the presence of the nontrivial solution is controlled solely by $G > G_\crit$,  
which is the same criticality condition as in  Eq.(\ref{crticality-condition}). 
The Lagrangian Eq.(\ref{with-explicit-breaking-0}) is also rewritten as 
\beq
{\cal L}^h 
&=& 
- G' 
\left( \bar{\chi}_L \chi_R - \vev{\bar{\chi}_L \chi_R }\right) 
\left( \bar{\chi}_R \chi_L - \vev{\bar{\chi}_R\chi_L}\right)
\,.\label{with-explicit-breaking}
\eeq
This implies that the explicit breaking effects can also be expressed only by the four-fermion interaction. 
We can thus approximately keep the global $U(3)_{\tilde{\psi}_L} \times U(1)_{\chi_R}$ invariance 
by taking $G'$ perturbatively to be 
\beq 
0< \frac{G'}{G} \ll 1 
\quad \text{and} \quad
G_\crit < G  
 \,. \label{G-prime}
\eeq
We have explicitly checked that in the presence of  this explicit breaking the vacuum 
with $m_{t\chi} \neq 0$ is preferred to that with $m_{t\chi} = 0$ 
in the phenomenologically interesting parameter space to be discussed later.

As in Sec.~\ref{Model-DSB}, 
the global $U(3)_{\tilde{\psi}_L}  \times U(1)_{\chi_R}$ symmetry is spontaneously broken down to 
$U(2)_{\tilde{\psi}_L}  \times U(1)_{V=\tilde{\chi}_L + \chi_R}$ 
due to the four-fermion interaction in ${\cal L}^{4f}$, 
resulting in the presence of five NGBs $\pi^a(a=4,5,6,7,A)$. 
Then the explicit breaking terms in ${\cal L}^{h}$ force the vacuum to choose a specific direction,  
$\vev{\bar{\chi}_R \tilde{\chi}_L} \neq 0$, and give masses to some of the NGBs. 
Note that the ${\cal L}^h$ term is invariant under the chiral transformation associated 
with the broken currents $(J_{3L}^{6\,,\mu} \pm i J_{3L}^{7\,,\mu})$ 
and $(J_{3L}^{4\,,\mu} \cos\theta + J_{3L}^{A\,,\mu} \sin\theta)$, 
but not for $J_{\mu 3L}^{5}$ and $(- J_{3L}^{4\,,\mu} \sin\theta + J_{3L}^{A\,,\mu} \cos\theta)$. 
Hence ${\cal L}^h$ term gives masses only to the NGBs associate with latter two, 
i.e., the Top-Mode Pseudos $A^0_t$ and $h^0_t$. 

Estimation  of the masses of the pseudo NGBs can be done by 
the traditional approach based on the current algebra~\cite{Dashen:1969eg}: 
\beq
m^2_{ab}
=
\frac{1}{f^2}
\left \langle 0 \left |
\left[ i Q^a \,, \left[ iQ^b\,,\,-{\cal L}^h \right] \right]
\right| 0 \right \rangle  
\,,\label{Dashen-formula}
\eeq
where $Q^a$ is the Noether's charge associated with 
the broken currents Eqs.(\ref{broken-current-3L4}), (\ref{broken-current-3L5}), 
(\ref{broken-current-3L6}), (\ref{broken-current-3L7}) and (\ref{broken-current-A}), 
and $f$ is given by Eq.(\ref{model-PSformula}). 
From Eq.(\ref{Dashen-formula}), together with the gap equations, 
Eqs.(\ref{gapeq-tchi}), (\ref{gapeq-chichi}) and (\ref{relation-Gapeq}), we obtain
\beq
m^2_{z^0_t} &=& m^2_{w^\pm_t} = 0
\label{mass-NGB}
\,,\\
m^2_{A^0_t} 
& =&
\frac{2 \vev{\bar{\chi}_R\tilde{\chi}_L} \vev{\bar{\chi}_R \chi_L}}{f^2\cos \theta} 
\nonumber\\
& \simeq &
\frac{G'}{G^2} \times \frac{2(m^2_{t\chi} + m^2_{\chi\chi})}{f^2} 
\,,\label{mass-At0}
\\[1ex]
m^2_{h^0_t} 
& =& 
\frac{2 \vev{\bar{\chi}_R\tilde{\chi}_L} \vev{\bar{\chi}_R \chi_L}}{f^2 \cos \theta}\times \sin^2 \theta\nonumber\\[1ex]
&=& 
m^2_{A^0_t} \sin^2\theta
\,,
\label{mass-ht0}
\eeq
where the second equation of Eq.(\ref{mass-At0}) is obtained by expanding in terms of $G'/G \ll 1$ 
and taking the leading nontrivial order. 
The mass of $h^0_t$ is proportional to $m_{t\chi}$ associated with 
the EWSB scale $v_{_{\rm EW}}$ as in  Eq.(\ref{def-vEW}), 
just like the case of the SM Higgs boson, while the mass of $A^0_t$ is not. 
We may set the mass of $h^0_t$ to $\simeq 126\,\GeV$;  
\beq
m_{h^0_t} = m_{A^0_t} \sin\theta \simeq 126\,\GeV
\,. \label{ht-At-mass-relation}
\eeq 
Note also that $G'>0$ assumed in Eq.(\ref{G-prime}) ensures 
the positiveness of squared masses 
for the Top-Mode Pseudos, $m^2_{h^0_t}>0$ and $m^2_{A^0_t} >0$.  
In Appendix.~\ref{app-BHL}, 
we present an alternative derivation of the pseudo NGBs masses 
and the heavy Higgs mass $m_{H^0_t}$ based on the approach used in~\cite{Bardeen:1989ds}. 

As was done in Eq.(\ref{NLsM-Lag-Op2}), 
we may construct a nonlinear Lagrangian valid for scales below $m_{H^0_t}$ 
described by the five NGBs 
based on the coset space in Eq.(\ref{TMP-coset}) 
including the explicit breaking effect 
from the $G'$- and $\Delta_{\chi\chi}$-terms in Eq.(\ref{with-explicit-breaking-0}). 
To this end, 
we introduce the spurion fields $\chi_1$ and $\chi_2$ 
to write the ${\cal O}(p^2)$ potential terms 
corresponding to Eq.(\ref{with-explicit-breaking-0}):  
\beq 
\Delta {\cal L}_{\text{NL$\sigma$M}}
= 
f^2 
 \tr
\left[
c_1 (R^T U)^\dagger \chi_1 (R^T U)
+ 
c_2 \left( \chi^\dagger_2 (R^T U) + (R^T U)^\dagger \chi_2 \right)
\right]
\,,
\label{NLsM-Lag-Op2-mass}
\eeq
where $\chi_1$ and $\chi_2$ transform under the ${\cal G}$-symmetry as 
\beq
\chi_1 \to g_{\tilde{3L}} \cdot \chi_1 \cdot g^\dagger_{\tilde{3L}} 
\,,\quad 
\chi_2 \to g_{\tilde{3L}} \cdot \chi_2 \cdot g^\dagger_{1R} 
\,.  
\eeq
The ${\cal G}$-symmetry is explicitly broken 
when the spurion fields acquire the vacuum expectation values, 
\beq
\vev{\chi_1} = \vev{\chi_2} = \Sigma 
\,, 
\eeq
so that the $c_1$ and $c_2$ terms break the ${\cal G}$ down to 
$U(2)_{q_L} \times U(1)_{\chi_L} \times U(1)_{\chi_R}$ 
and $U(2)_{q_L} \times U(1)_{V=\chi_R+\chi_L}$, respectively, 
in the same way as the $G'$ and $\Delta_{\chi\chi}$ terms 
in Eq.(\ref{with-explicit-breaking-0}) do. 
Matching to the tHiggs mass formula in Eq.(\ref{mass-ht0}), 
we then find the coefficients $c_1$ and $c_2$ to be 
\beq
c_1 = - \frac{1}{2} m^2_{A^0_t} 
\,, \qquad 
c_2 = \frac{1}{2} m^2_{A^0_t} \cos \theta 
\,.\nonumber
\eeq

\subsection{Fermion masses and Yukawa couplings}
\label{fermion-mass}

\subsubsection{top and $t'$-quark}

Let us consider the top quark mass based on the Lagrangian Eq.(\ref{with-explicit-breaking-00}). 
After the spontaneous symmetry breaking by the nontrivial solutions of the gap equations, 
Eqs.(\ref{gapeq-tchi}), (\ref{gapeq-chichi}) and (\ref{relation-Gapeq}), 
the mass terms of the top quark $t$ and its flavor partner $\chi$ look like
\beq
\left.
{\cal L}_{\rm kin.} +{\cal L}^{4f}+ {\cal L}^h\right|_{\rm mass}
=
-\bpm \bar{t}_L & \bar{\chi}_L \epm
\bpm 0 & m_{t\chi} \\ 0 & m_{\chi \chi} \epm
\bpm t_R \\ \chi_R \epm + \text{h.c.}
\,.\label{f-mass-matrix}
\eeq
From this we find the fermion mass eigenvalues as 
\beq
m^2_{t'} = m^2_{\chi\chi} + m^2_{t\chi}
\quad , \quad
m^2_t = 0
\,,\label{massless-top}
\eeq
where the mass eigenstates $t'$ and $t$ are given by 
\beq
t'_L = \tilde{\chi}_L = t_L \sin \theta + \chi_L \cos \theta 
\quad , \quad
t_L = \tilde{t}_L = t_L \cos \theta - \chi_L \sin \theta
\,.\nonumber
\eeq
Thus the top quark does not ``feel" the EWSB by $\sin\theta \neq 0 \, (m_{t\chi} \neq 0)$ 
and is still massless. 
This is essentially due to the residual symmetry, 
$U(2)_{q_R}  \times U(1)_{V=\tilde{\chi}_L + \chi_R}$ ($U(2)_{q_R}: t_R \leftrightarrow b_R$), 
which forbids the couplings between $t_R$ and $\chi_L$ 
in the Lagrangian Eq.(\ref{with-explicit-breaking-00}), 
hence no mass term for $\bar{\chi}_L t_R$ in Eq.(\ref{f-mass-matrix}).   

To make the model more realistic, 
we introduce a four-fermion interaction term which breaks the residual symmetry 
so as to allow $t_R$ to couple to $\chi_L$:    
\beq
{\cal L}_{\rm kin.} +
{\cal L}^{4f}
+ {\cal L}^h
+ {\cal L}^t
\,,   \label{with-top-mass}
\eeq
where
\beq 
{\cal L}^t
=
G'' \left( \bar{\chi}_L \chi_R \right) \left( \bar{t}_R \chi_L\right) + \text{h.c.} 
\,. \label{4f-top-mass}
\eeq
As was done in Eq.(\ref{G-prime}), we also treat the $G''$ coupling to be perturbative, 
\beq 
0< \frac{G''}{G} < 1 
\,, \label{G-double-prime}
\eeq
so that the symmetry breaking pattern 
${\cal G}/{\cal H}
= [U(3)_{\tilde{\psi}_L} \times U(1)_{\chi_R}]/[U(2)_{\tilde{\psi}_L} \times U(1)_{V}]$ 
is not destroyed 
(the vacuum aligned to this manifold): 
$\vev{\bar{t}_R \chi_L}_{G''=0} = \vev{\bar{t}_R t_L}_{G''=0} =0$:   
The Dashen formula Eq.(\ref{Dashen-formula}) with the $G''$ term in Eq.(\ref{4f-top-mass}) 
leads to 
\beq
\left. m^2_{h^0_t} \right|_{G''}
&=&
\frac{1}{f^2}
\left \langle 0 \left |
\left[ i Q^5 \,, \left[ iQ^5 \,,\,  - {\cal L}^t \right] \right]
\right| 0 \right \rangle  
\nonumber\\ 
 &=& 
\frac{G''}{f^2} \left[
(\cos^2\theta - \sin^2\theta)
\vev{\bar{\tilde{\chi}}_L t_R}\vev{\bar{t}_R \tilde{\chi}_L }
+
2\sin\theta\cos\theta
\vev{\bar{\tilde{\chi}}_L \chi_R}\vev{\bar{t}_R \tilde{t}_L}
\right] 
\nonumber \\ 
&\simeq & 
G''\left[
\vev{ \bar{\chi}_R \chi_L}_{G''=0} \vev{\bar{t}_R \chi_L}_{G''=0} 
-
\vev{\bar{\chi}_R t_L}_{G''=0} \vev{\bar{t}_R t_L}_{G''=0} 
\right]
+ {\cal O}((G'')^2) 
\nonumber\\[1ex]
&=& 
0 +  {\cal O}((G'')^2) 
\,, \label{Dashen-Formula-G''}
\eeq 
similarly for the mass of the CP-odd Top-Mode Pseudo, $A^0_t$. 
Here we have granted the vacuum saturation valid at the leading order of $1/N_c$. 
Thus the Top-Mode Pseudo's masses are stable against the leading order correction 
of the explicit-breaking $G''$ term as dictated by the Dashen formula Eq.(\ref{Dashen-Formula-G''}).
The next-to leading order in the $G''$-perturbation, i.e., $(G''/G)^2$ corrections 
(which are also of the leading order of $1/N_c$), 
will affect the masses, 
as will be discussed later (see around Eq.(\ref{quadratic-divergence-mh})). 

After the spontaneous symmetry breaking by $\vev{\bar{\chi}_R \tilde{\chi}_L} \neq 0$, 
i.e. $m^2_{\tilde{\chi} \chi} = m^2_{t\chi} + m^2_{\chi\chi} \neq 0$, 
we thus have the fermion mass matrix,  
\beq
\left.
{\cal L}_{\rm kin.} +
{\cal L}^{4f}
+ {\cal L}^h
+ {\cal L}^t
\right|_{\rm mass}
=
-\bpm \bar{t}_L & \bar{\chi}_L \epm
\bpm 0 & m_{t\chi} \\ \mu_{\chi t} & m_{\chi \chi} \epm
\bpm t_R \\ \chi_R \epm + \text{h.c.}
\,,\label{mass-ttprime}
\eeq
where $\mu_{\chi t}= - G'' \vev{\bar{\chi}_R \chi_L} $ is a dynamical mass coming from ${\cal L}^t$ 
in Eq.(\ref{with-top-mass}). 
Note that the fermion mass matrix is identical to that discussed 
in top-seesaw models~\cite{Dobrescu:1997nm,Chivukula:1998wd}. 
The top quark and $t'$-quark masses are given as the eigenvalues of the mass matrix 
in Eq.(\ref{mass-ttprime}), 
\beq
m^2_{t'} 
&=& 
\frac{m^2_{t\chi}+m^2_{\chi\chi}+\mu^2_{\chi t}}{2}
\left[
1+
\sqrt{
1-\frac{4m^2_{t\chi}\mu^2_{\chi t}}{\left( m^2_{t\chi}+m^2_{\chi\chi}+\mu^2_{\chi t} \right)^2}
}
\right]
\,,\label{Model-t'mass}
\\[1ex] 
m^2_t 
&=& 
\frac{m^2_{t\chi}+m^2_{\chi\chi}+\mu^2_{\chi t}}{2}
\left[
1-
\sqrt{
1-\frac{4m^2_{t\chi}\mu^2_{\chi t}}{\left( m^2_{t\chi}+m^2_{\chi\chi}+\mu^2_{\chi t}\right)^2}
}
\right]
\,.\label{Model-tmass}
\eeq
Now the top quark mass becomes nonzero and is proportional to $m_{t\chi}$ 
which breaks the electroweak symmetry, similarly to the mass of $h^0_t$ in Eq.(\ref{mass-ht0}), 
while the $t'$-quark mass is not. 
The corresponding mass eigenstates $(t,t')^T_m$ are related to 
the gauge (current) eigenstates $(t,\chi)^T_g$ by the orthogonal rotation 
keeping $m_t,m_{t'} \geq 0$~\cite{He:2001fz}: 
\beq
\bpm t_{L} \\[1ex] t'_{L} \epm_m
=
\bpm c^t_{L} & -s^t_{L} \\[1ex] s^t_{L} & c^t_{L}\epm
\bpm t_{L} \\[1ex] \chi_{L} \epm_g
=
O_L \cdot \bpm t_{L} \\[1ex] \chi_{L} \epm_g
\quad , \quad
\bpm t_{R} \\[1ex] t'_{R} \epm_m
=
\bpm -c^t_{R} & s^t_{R} \\[1ex] s^t_{R} & c^t_{R}\epm
\bpm t_{R} \\[1ex] \chi_{R} \epm_g
=
O_R \cdot \bpm t_{R} \\[1ex] \chi_{R} \epm_g
\,,\label{rotate-ttprime}
\eeq
where $c^t_{L(R)} \equiv \cos \theta^t_{L(R)}$ and $s^t_{L(R)} \equiv \sin \theta^t_{L(R)}$ 
which are given up to ${\cal O}((G''/G)^2)$ 
as  
\beq 
c^t_L 
&=& 
\frac{1}{\sqrt{2}} \left[ 
1 + \frac{m^2_{\chi \chi} -m^2_{t\chi} + \mu^2_{\chi t}}{m^2_{t'} - m^2_t}
\right]^{1/2}
\simeq
\cos \theta \left[
1+ \left( \frac{G''}{G} \right)^2 \cos^2\theta \sin^2\theta
\right]
\,,\label{def-cLt}
\\
s^t_L 
&=& 
\frac{1}{\sqrt{2}} \left[ 
1 - \frac{m^2_{\chi \chi} -m^2_{t\chi} + \mu^2_{\chi t}}{m^2_{t'} - m^2_t}
\right]^{1/2}
\simeq
\sin \theta \left[ 
1-\left( \frac{G''}{G} \right)^2 \cos^4\theta 
\right]
\,,\label{def-sLt}
\\
c^t_R
&=& 
\frac{1}{\sqrt{2}} \left[ 
1 + \frac{m^2_{\chi \chi} + m^2_{t \chi}  -\mu^2_{\chi t}}{m^2_{t'} - m^2_t}
\right]^{1/2}
\simeq
1 - \frac{1}{2}\left( \frac{G''}{G} \right)^2 \cos^4\theta
\,,\label{def-cRt}
\\
s^t_R
&=& 
\frac{1}{\sqrt{2}} \left[ 
1 - \frac{m^2_{\chi \chi} + m^2_{t \chi}  -\mu^2_{\chi t}}{m^2_{t'} - m^2_t}
\right]^{1/2}
\simeq
\frac{G''}{G}\cos^2\theta 
\,.\label{def-sRt}
\eeq

Including the effect of the explicit breaking $G''$ term, 
we may thus add the fermion sector to the nonlinear Lagrangian 
constructed from Eqs.(\ref{NLsM-Lag-Op2}) and (\ref{NLsM-Lag-Op2-mass}), 
\beq
{\cal L}^{t,t'}_{\rm yuk.}
=
-
\frac{f}{\sqrt{2}}
\left[ 
y \bar{\psi}_L (R^T U) \psi_R 
+
y_{\chi t}\bar{\psi}_L (\chi_1 R^T U \chi_3) \psi_R
+ \text{h.c.}
\right]
\,,  
\label{NLsM-Lag-Op2-withtopmass}
\eeq
where $\psi_R=(t_R, b_R, \chi_R)^T$. 
The spurion fields $\chi_1$ and $\chi_3$ have been introduced 
in Eq.(\ref{NLsM-Lag-Op2-withtopmass}), 
which transform as 
\beq
\chi_1 \to g_{3 L} \cdot \chi_1 \cdot g^\dagger_{3L} 
\,, \quad
\chi_3\to g_{1R} \cdot \chi_3 \cdot g^\dagger_{1R}
\,,  
\eeq
so that the Lagrangian Eq.(\ref{NLsM-Lag-Op2-withtopmass}) is invariant 
under the ${\cal G}$-symmetry, 
$U(3)_{\tilde{\psi}_L} \times U(1)_{\chi_R}$, and $U(2)_{q_R}$ symmetry. 
These symmetries are explicitly broken by the vacuum expectation values of the spurion fields, 
\beq 
\vev{\chi_1} = \Sigma 
\,,\quad 
\vev{\chi_3} = \lambda_4 
\,, 
\eeq
in which the $\vev{\chi_1}$ breaks the $U(3)_{\psi_L}$ symmetry down to 
$U(2)_{\psi_L} \times U(1)_{\chi_L}$ 
and the $\vev{\chi_3}$ does the $U(2)_{q_R}\times U(1)_{\chi_R}$ down to 
$U(1)_{\chi_R=t_R}$. 
We thus see that the first term in Eq.(\ref{NLsM-Lag-Op2-withtopmass}) 
corresponds to the $G$-four fermion term in Eq.(\ref{start-Lag}), 
which is invariant under the ${\cal G}$-symmetry, 
$U(3)_{\tilde{\psi}_L} \times U(1)_{\chi_R}$ and $U(2)_{q_R}$ symmetry,  
while the second term does to the $G''$-four fermion term, 
which explicitly breaks the ${\cal G}$ and $U(2)_{q_R}$ down to 
$U(2)_{q_L} \times U(1)_{\chi_L}$ and $U(1)_{t_R=\chi_R}$. 
The Yukawa couplings $y$ and $y_{\chi t}$ can be
fixed by matching to the fermion mass matrix in Eq.(\ref{mass-ttprime}) as  
\beq
y^2 = \frac{2(m^2_{t\chi}+m^2_{\chi\chi})}{f^2} 
\quad , \quad
y^2_{\chi t} 
= 
y^2 \left( \frac{G''}{G}\right)^2
= 
\frac{2\mu^2_{\chi t}}{f^2 \cos^2\theta}
\,.\label{def-yukawa}
\eeq

\subsubsection{Fermions other than top and $t'$-quark}

In order to give masses to SM fermions other than the top quark, 
inspired by~\cite{Miransky:1988xi,Miransky:1989ds}, 
we may add the following four-fermion interactions to the Lagrangian Eq.(\ref{with-top-mass}): 
\beq
{\cal L}^{\rm others}
=
\sum_{\alpha=1,2} 
G^\alpha_{tu} (\bar{q}^i_L \chi_R) (\bar{u}^{\alpha}_R q^{\alpha,i}_L)
+
\sum_{\alpha=1,2,3} 
G^\alpha_{td} (\bar{q}^i_L \chi_R)(i\tau^2)^{ij}(\bar{q}^{\alpha,j}_L d^{\alpha}_R) 
+ 
\sum_{\alpha=1,2,3} 
G^\alpha_{te} (\bar{q}^i_L \chi_R)(i\tau^2)^{ij}(\bar{l}^{\alpha,j}_L e^{\alpha}_R) 
+ \text{h.c}
\,,\label{btau-mass}
\eeq
which are SM gauge invariant, 
where $\alpha=1,2,3$ denotes the index for the fermion generation 
and $i,j=1,2$ for the weak isospin. 
With the nonzero condensate $\vev{\bar{\chi}_R t_L} \neq 0$ ($m_{t\chi} \neq 0$) breaking 
the electroweak gauge  symmetry, 
these SM fermions acquire their masses as 
\beq
m_{u^\alpha} = - G^\alpha_{tu} \vev{\bar{\chi}_R t_L}
\quad , \quad
m_{d^\alpha} = -G^\alpha_{td} \vev{\bar{\chi}_Rt_L}
\quad , \quad
m_{e^\alpha} = -G^\alpha_{te} \vev{\bar{\chi}_Rt_L}
\,.\nonumber
\eeq 

Since the NGBs arise from $\bar{\chi}_R \psi_L$ as
\beq
\bar{\chi}_R q^i_L
\sim
\vev{\bar{\chi}_R \tilde{\chi}_L}  [ R^T U]_{3i}
\,,
\eeq
one can read off the fermion couplings to $h^0_t$~\footnote{
Those Yukawa interactions would give quadratically divergent corrections to the $h^0_t$ mass, 
which are, however, small enough due to the small Yukawa coupling for the lighter fermions, 
to be negligible compared to the terms in Eq.(\ref{quadratic-divergence-mh}) arising from 
the Top-Mode Pseudos, $t$ and $t'$-loops. 
} :
\beq
\left.
{\cal L}^{\rm others}_{\rm yuk.}
\right|_{h^0_t}
=
- \cos \theta \left[ 
\sum_{\alpha=1,2} \frac{m_{u^\alpha}}{v_{\rm EW}} h^0_t\bar{u}^\alpha u^\alpha
+
\sum_{\alpha=1,2,3}\frac{m_{d^\alpha}}{v_{\rm EW}} h^0_t\bar{d}^\alpha d^\alpha
+
\sum_{\alpha=1,2,3}\frac{m_{e^\alpha}}{v_{\rm EW}} h^0_t\bar{e}^\alpha e^\alpha
\right]
\,. \label{yukawa-no-top}
\eeq 
A full set of the particle content in the present model with the SM charge assignment free from 
the gauge anomaly is listed in Table.~\ref{contents-model}. 

\begin{table}[h]
\begin{center}
\begin{tabular}{| c ||  c | c | c |}
\hline
field  & $SU(3)_c$ & $SU(2)_L$ & $U(1)_Y$ 
\\ 
\hline
\, & \, & \, & \,  
\\[-2.5ex] 
\hline
$q_L= \bpm t_L \\ b_L \epm$ & 3 & 2 & 1/6   
\\
$t_R$ & 3 & 1 & 2/3 
\\
$b_R$ & 3 & 1 & -1/3 
\\
\hline 
\, & \, & \, & \, 
\\[-2.5ex] 
\hline 
$l_L = \bpm \nu_{\tau L} \\ \tau_L \epm$ & 1 & 2 & -1/2  
\\ 
$\tau_R$ & 1 & 1 & -1 
\\ 
\hline 
\, & \, & \, &\,
\\[-2.5ex]
\hline
$\chi_L$ & 3 & 1 & 2/3  
\\
$\chi_R$ & 3 & 1 & 2/3  
\\
\hline
\end{tabular}
\caption{Full particle contents and the charge assignments under the SM gauge group. 
All the fermions are represented in terms of the electroweak gauge eigenbasis. } 
\label{contents-model}
\end{center}
\end{table}%

\subsection{tHiggs}
\label{TNH}

We here discuss the coupling property of the tHiggs $h^0_t$ 
and the stability of the mass against radiative corrections. 
From Eqs.(\ref{NLsM-Lag-Op2}), (\ref{NLsM-Lag-Op2-mass}), 
(\ref{NLsM-Lag-Op2-withtopmass}) and (\ref{yukawa-no-top}), 
we find the $h^0_t$ couplings to the SM particles, 
\beq
\left.
{\cal L}_{\text{NL$\sigma$M}}+
\Delta {\cal L}_{\text{NL$\sigma$M}}+
{\cal L}^{t,t'}_{\rm yuk.} + 
{\cal L}^{\rm others}_{\rm yuk.}
\right|_{h^0_t}
&=&
g_{hVV} \frac{v_{_{\rm EW}}}{2}  
\left( g^2 h^0_t W^+_\mu W^{-\mu} 
+ 
\frac{g^2 + g'^2}{2} h^0_t Z_\mu Z^\mu \right)
\nonumber\\
&&
-
g_{hhh}\frac{3m^2_{h^0_t}}{v_{_{\rm EW}}} \frac{\left( h^0_t \right)^3}{3!} 
-
g_{hhhh}\frac{3m^2_{h^0_t}}{v^2_{_{\rm EW}}}
\frac{\left( h^0_t \right)^4}{4!}
\nonumber\\
&&
-g_{htt} \frac{m_t}{v_{_{\rm EW}}} h^0_t \bar{t}t 
-g_{hbb} \frac{m_b}{v_{_{\rm EW}}} h^0_t \bar{b}b 
-g_{h\tau\tau} \frac{m_\tau}{v_{_{\rm EW}}} h^0_t\bar{\tau}\tau
\nonumber\\
&&
+ \cdots
\,,\label{Model-higgs-coupling}
\eeq
where 
\beq
&&
g_{hVV} 
= g_{hhh}
= g_{hbb} 
= g_{h\tau\tau}
= \cos \theta
\,,\label{gTM-hVVhhff}\\
&&
g_{hhhh} = 1 - \frac{7}{3} \sin^2\theta
\,,\label{gTM-hhhh}\\
&&
g_{htt} 
=
\frac{v_{_{\rm EW}}}{m_t}\frac{y}{\sqrt{2}} 
\left[ 
(c^t_L \cos \theta  + s^t_L \sin \theta ) s^t_R 
- 
s^t_Lc^t_R \sin\theta \left( \frac{G''}{G} \right)
\right]
= 
\frac{2\cos^2\theta -1}{\cos\theta} + {\cal O} \left( \left(\frac{G''}{G}\right)^2\right)
\,.\label{gTM-htt}
\eeq
From these, 
we see that the $h^0_t$ couplings to the $W,Z$ bosons and to the SM fermions 
become the same as the SM Higgs ones 
when we take the limit $\cos \theta \to 1$, i.e., 
\beq
g_{hVV} = g_{hhh} = g_{hhhh} = g_{htt} = g_{hbb} = g_{h\tau\tau} = g^{_{\rm SM}} (= 1)
\,,\nonumber
\eeq
when
\beq
\sin\theta 
= 
\frac{m_{t\chi}}{\sqrt{m_{t\chi}^2 + m_{\chi\chi}^2}}
= 
\frac{v_{_{\rm EW}}}{f} \to 0 
\,, \qquad \text{by} \quad 
f \to \infty \quad \text{with} \quad v_{_{\rm EW}} = 246\, \GeV \quad \text{fixed}
\,. \label{SMlimit}
\eeq
Actually, this limit turns out to be favored by the current experiments 
as will be discussed in Sec.~\ref{phenomenology-TMP}.

From Eqs.(\ref{NLsM-Lag-Op2}), (\ref{NLsM-Lag-Op2-mass}), 
(\ref{NLsM-Lag-Op2-withtopmass}) and (\ref{yukawa-no-top}), 
we can also evaluate the quadratic divergent corrections to the $h^0_t$ mass 
at the one-loop level. 
The one-loop corrections of the leading order of the explicit breaking parameters 
$G'/G\,,G''/G$ and $\alpha_{\rm em}=e^2/(4\pi)$ 
with $e$ being the electromagnetic coupling, 
can be evaluated at the one-loop level of the 
nonlinear sigma model constructed from terms in 
Eqs.(\ref{NLsM-Lag-Op2}), (\ref{NLsM-Lag-Op2-mass}), 
(\ref{NLsM-Lag-Op2-withtopmass}) and (\ref{yukawa-no-top}).  
The correction of ${\cal O}(G''/G)$ from $t$ and $t'$ loops exactly cancel 
and do not contribute to the tHiggs mass, 
which is consistent with the Dashen formula in Eq. (\ref{Dashen-Formula-G''}). 
Hence the leading order corrections in the perturbation 
with respect to the explicit breaking couplings 
only come from terms of ${\cal O}(G'/G)={\cal O}(m^2_{h^0_t})$ and ${\cal O}(\alpha_{\rm em})$, 
in which the quadratic divergent contributions dominate. 
Of these leading corrections, 
the electroweak gauge terms are actually highly suppressed by $\alpha_{\rm em}$ 
compared to the corrections of ${\cal O}((G'/G)\Lambda^2_\chi/(4\pi)^2)$ 
which arises from the Top-Mode Pseudo's ($h^0_t$ and $A^0_t$) self-interaction sector, 
where $\Lambda_\chi \sim m_{H^0_t}$ is the cutoff of the nonlinear sigma model. 
Thus we evaluate the leading order corrections 
in the perturbation of $G'/G\,,G''/G$ and $\alpha_{\rm em}$ to the tHiggs mass: 
\beq
\left. m^2_{h^0_t} \right|^{{\cal O}(G'/G, G''/G, \alpha_{\rm em})}_{\rm 1-loop}
=
m^2_{h^0_t}
\left[
1 + 
\frac{\Lambda_\chi^2}{(4\pi)^2v^2_{_{\rm EW}}} 
 \frac{23}{16} 
 \right]
+ 
{\cal O}
\left(
\frac{\alpha_{\rm em}\Lambda^2_\chi}{4\pi} 
\right) 
\,.\label{quadratic-divergence-mh:meson-loop}
\eeq
The size of $G'/G$ correction 
(the second term in the square bracket) 
could be ${\cal O}(1)$, 
when we took $\Lambda_\chi \sim m_{H^0_t} \sim 4 \pi v_{_{\rm EW}}$, 
and hence is potentially a large correction to the tHiggs mass, 
which might need some fine tuning. %

Actually, the top and $t'$-loop corrections arising 
as the next to leading order of ${\cal O}((G''/G)^2)$ will be more significant 
to give the sizable contribution to the tHiggs mass at the one-loop order 
since $G''/G$ is numerically not very small in order to realize the reality, 
though those terms are potentially suppressed in terms of the $(G''/G)$-perturbation. 
The concrete estimate of the size of corrections to the tHiggs mass 
from those terms will be addressed in Sec.~\ref{summary}. 

One might naively suspect from Eqs.(\ref{quadratic-divergence-mh:meson-loop}) that 
the presence of quadratic divergent corrections to the $h^0_t$ mass causes 
the fine-tuning problem just like the SM Higgs boson case. 
However, it is not the case because the $h^0_t$ is \textit{natural} 
in accordance with the original argument in~\cite{'tHooft:1979bh}: 
If one takes the massless Higgs boson limit ($m_{h^0_t}  \to 0$) corresponding to 
$G' \to 0\,,G'' \to 0\,\,
( \text{and}\,\,G^\alpha_{tu} \to 0\,,G^\alpha_{td} \to 0\,,G^\alpha_{te} \to 0)$
and  electroweak gauge interactions are turned off 
in Eqs.(\ref{NLsM-Lag-Op2}), (\ref{NLsM-Lag-Op2-mass}) and (\ref{NLsM-Lag-Op2-withtopmass}) 
(and Eq.(\ref{yukawa-no-top})), 
then the global ${\cal G}$-symmetry is restored, 
meaning that the symmetry is enhanced. 
In this limit the quadratic divergences disappear as well. 
Thus the $h^0_t$ mass is protected by the ${\cal G}$-symmetry 
just like the QCD pseudo NGBs ($\pi, K \cdots$). 
In contrast, as is well known, the SM Higgs sector is \textit{unnatural} 
since even if one takes the massless Higgs boson limit ($m_{h^0_{_{\rm SM}}}  \to 0$) 
the symmetry of the SM is not enhanced. 
In this sense, the $h^0_t$ is a natural Higgs.

Before closing this section, 
it is also worth mentioning the difference between the present model and 
little Higgs models~\cite{ArkaniHamed:2002qx,ArkaniHamed:2002qy,Cheng:2003ju,Cheng:2004yc}. 
They are similar in the sense that one of NGBs associated with 
the spontaneous global symmetry breaking (${\cal G}/{\cal H}$) is identified as the Higgs boson, 
and the Higgs mass is generated through the explicit breaking effect. 
Therefore, the Higgs mass is under control in both models as explained in the previous paragraph. 
The crucial difference, though, is that, in the case of little Higgs models, 
the electroweak symmetry is embedded as a subgroup of ${\cal H}$, 
while in the case of the present model, it is outside of ${\cal H}$, namely, 
the electroweak symmetry is broken by the dynamics which breaks ${\cal G}$ down to ${\cal H}$.

\section{Phenomenologies}
\label{phenomenology}

In this section, we discuss several constraints from existing experimental results 
and phenomenological implications of the model.

\subsection{\textit{Phenomenological constraints on Top-Mode Pseudos}}
\label{phenomenology-TMP}

Examining Eqs.(\ref{Model-higgs-coupling}) and (\ref{gTM-hVVhhff}), 
we see that the couplings of the tHiggs  $h^0_t$ to the $W$ and $Z$ bosons 
deviate from the SM Higgs ones by  
\beq
\kappa_V 
\equiv
\frac{g_{hVV}}{g^{_{\rm SM}}_{hVV}} = \cos \theta
\,,\label{hVV-ratio}
\eeq
where $V=W$ and $Z$. 
The current LHC data give the constraint on $\kappa_V$ to be 
$\kappa_V > 0.94$ at $95\%\cl$ for the 126 GeV Higgs boson~\cite{ATLAS:2013sla}. 
Therefore, we obtain the following constraint on the angle $\theta$:
\beq
\cos \theta > 0.94, \quad \sin \theta < 0.34.
\label{bound-theta}
\eeq

As noted in Eq.(\ref{SMlimit}), 
the tHiggs couplings to the SM particles coincide with the SM Higgs ones when $\theta \ll 1$: 
The deviation in the ratio of the couplings to fermions to those of the $W$ and $Z$ bosons, 
$g_{hff}/g_{hVV}$, can be expanded in powers of $\theta$ as 
\beq 
\frac{g_{h ff}}{g_{hVV}} &=& 1\,, \qquad {\rm for} \qquad f=\tau, b,
\nonumber \\  
\frac{g_{htt}}{g_{hVV}} &\simeq& 1 - \frac{3}{2}\theta^2 
\,. \nonumber
\eeq
This and the current bound on $\theta$ in Eq.(\ref{bound-theta}) imply that 
the highly precise measurement (by about $5\%$ accuracy) 
would be required to distinguish 
the coupling properties between the SM Higgs and the tHiggs. 
It might be possible to make it by the high luminosity LHC or ILC~\cite{Peskin:2012we}. 

Here, it is also worth giving some comments on the difference 
between the present low-energy effective theory 
(Eqs.(\ref{NLsM-Lag-Op2}),(\ref{NLsM-Lag-Op2-mass}) and (\ref{NLsM-Lag-Op2-withtopmass})) 
and the two-Higgs doublet model 
since the top-seesaw model is often described by using a two Higgs doublet model 
as the low-energy effective theory~\cite{He:2001fz,Fukano:2012qx}. 
One difference is in the low-energy mass spectrum: 
As seen in Sec.\ref{TMPs}, 
the low-energy mass spectrum in the present model has no charged Higgs bosons 
which the two-Higgs doublet model posses.   
The other would be the tree-level mass relation among the neutral Higgs bosons 
($A^0_t$ and $h^0_t$) 
as in Eq.(\ref{ht-At-mass-relation}), 
which is absent in the usual two-Higgs doublet model 
and therefore may distinguish two models.

\subsection{\textit{$S$, $T$ parameters and the constraint on $t'$-quark mass}}

Looking at the current bound on $\theta$ in Eq.(\ref{bound-theta}), 
we see that the coupling property of the $t'$-quark arises mainly from 
the $SU(2)_L$ singlet $\chi$-quark which carries exactly the same charge as 
that of the right-handed top quark. 
The mass of $t'$-quark can therefore be constrained 
from the Peskin-Takeuchi $S,T$-parameters~\cite{Peskin:1990zt,Peskin:1991sw}~\footnote{
Another possible constraint would be $t'$-quark contribution to the $Zb_L\bar{b}_L$ coupling, 
which, however, turns out to be much milder than that from the $S,T$ parameters. 
This is due to the fact that the present model does not include $b'$-like particle usually arising 
in a class of top-seesaw models 
with so-called bottom-seesaw mechanism~\cite{He:2001fz,Fukano:2012qx}.
}. 

\begin{figure}[htbp]
\begin{center}
\begin{tabular}{cc}
{
\begin{minipage}[t]{0.4\textwidth}
\includegraphics[scale=0.6]{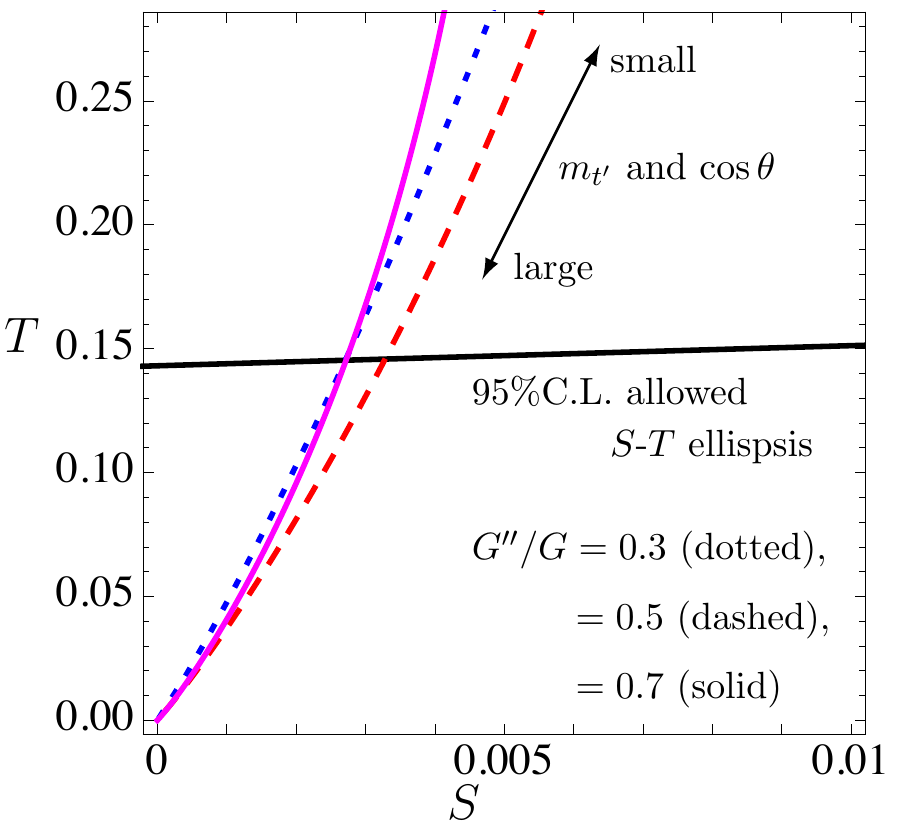} 
\end{minipage}
}
&
{
\begin{minipage}[t]{0.5\textwidth}
\includegraphics[scale=0.7]{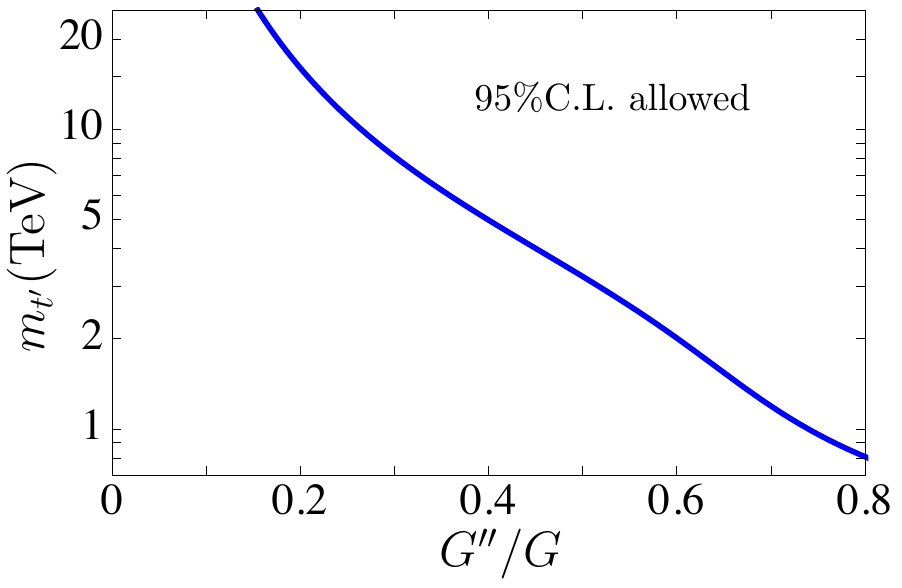} 
\end{minipage}
}
\end{tabular}
\caption[]{
Left panel: 
The $S,T$ constraints from Eqs.(\ref{present-S}) and (\ref{present-T}) on 
the $t'$-quark mass in the $(S,T)$-plane for $G''/G = 0.3$ (dotted), $0.5$ (dashed), $0.7$(solid). 
The $95\%\cl$ allowed region corresponds to an area 
lower than the solid curve (inside the $S$-$T$ ellipsis). 
The region of $S < 0$ and $T<0$ are not displayed 
due to $c^t_L \leq 1$~\footnotemark[1]. 
Right panel:  
The $t'$-quark mass versus $G''/G$ allowed by the $S,T$ constraints of the left panel. 
The $95\%\cl$ allowed region corresponds to an area upper than the blue curve.  
\label{ST-constraint}}
\end{center}
\end{figure}

By taking $m_{t'} \gg m_t \gg m_b$, 
the one-loop contributions from the $t'$-quark to 
the $S,T$-parameters are evaluated as~\cite{Chivukula:1998wd}
\beq
S
&=&
\frac{3}{2\pi} (s^t_L)^2 
\left[ 
-\frac{1}{9}\ln\frac{x_{t'}}{x_t} -  (c^t_L)^2 F (x_t,x_{t'}) 
\right]
\,,\label{present-S}
\\
T
&=& 
\frac{3}{16 \pi s^2_W c^2_W} (s^t_L)^2
\left[ 
(s^t_L)^2 x_{t'} 
- 
\left( 1+(c^t_L)^2 \right) x_t 
+ 
(c^t_L)^2 \frac{2x_{t'}x_t}{x_{t'}-x_t}\ln\frac{x_{t'}}{x_t}
\right]\,,
\label{present-T}
\eeq 
where $s_W \equiv\sin\theta_W (c^2_W \equiv 1-s^2_W)$ is the weak mixing angle 
and $x_a \equiv m^2_a/m^2_Z\,,(a=t,t')$. 
The function $F(x,y)$ is given by~\cite{Lavoura:1992np}
\beq
F(x,y) 
= 
\frac{5(x^2 + y^2)-22xy}{9(x-y)^2} 
+
\frac{3xy(x+y)-x^3-y^3}{3(x-y)^3} \ln\frac{x}{y}
\,.\nonumber
\eeq 
The $S,T$-parameters in Eqs.~(\ref{present-S}) and (\ref{present-T}) 
are calculated as a function of the two parameters, 
$s^t_L(c^t_L)$ and $m_{t'}$ 
once we fix $m_Z$, $s_W(c_W)$ and $m_t$ to be the experimental values~\cite{Beringer:1900zz}. 
Expanding Eqs.(\ref{Model-t'mass}) and (\ref{Model-tmass}) 
in powers of $G''/G = \mu_{\chi t}/m_{\chi\chi} <1$ 
we express the ratio $m_t/m_{t'}$ to the next to leading order of $G''/G$: 
\beq
\frac{m_{t}}{m_{t'}} 
\simeq 
\sin\theta \cos\theta \left( \frac{G''}{G} \right) 
\left[ 
1 - \cos^4\theta \left( \frac{G''}{G} \right) ^2
\right]
=  
\frac{\mu_{\chi t}}{m_{\chi\chi}} \sin\theta \cos\theta 
\left[ 
1 - \cos^4\theta \left( \frac{\mu_{\chi t}}{m_{\chi\chi}} \right) ^2
\right]
\,.
\eeq
Using  this and Eq.(\ref{def-sLt}), 
we can rewrite $s^t_L(c^t_L)$ in terms of $\theta$ and $G''/G$ 
to evaluate the $S$ and $T$ as a function of $\theta$ and $G''/G$. 
In Fig.~\ref{ST-constraint} (left panel), 
we thus plot the $S,T$-parameters versus $\cos\theta$ 
for several values of $G''/G$, 
together with $95\%\cl$ allowed region (inside of the ellipsis) 
on the $(S,T)$-plane~\cite{Ciuchini:2013pca}. 
We have taken into account the constraints on $\cos\theta$ in Eq.(\ref{bound-theta}) 
from the current LHC Higgs search. 
From the right panel of Fig.~\ref{ST-constraint}, we read off the allowed $t'$-quark mass, 
\beq
m_{t'} \geq 
\left\{
\begin{aligned}
& 8.11\,\TeV, 
\quad \quad
\cos \theta \geq 0.997
& \quad \text{for $\dfrac{G''}{G}=0.3$}
\,,
\\[1ex]
& 3.23\,\TeV, 
\quad \quad
\cos \theta \geq 0.991
& \quad \text{for $\dfrac{G''}{G}=0.5$}
\,,
\\[1ex]
& 1.19\,\TeV, 
\quad \quad
\cos \theta \geq 0.952
& \quad \text{for $\dfrac{G''}{G}=0.7$}
\,,
\end{aligned}
\right.
\label{ST-t'}
\eeq
where the lower mass limits corresponds to 
the $95 \%\cl$ upper limit of the $ST$-ellipsis  
in the left panel of Fig.~\ref{ST-constraint}. 
Note that the limit $\cos\theta \to 1\,(\sin\theta \to 0)$ corresponds to the decoupling limit of $t'$-quark, 
$m_{t'} \to \infty$ where $s^t_L \to 0$.
The stringent phenomenological constraints on the $t'$-quark 
thus come from the contribution to the $S,T$ parameters, 
which limit the mass to be $\gtrsim {\cal O}(\TeV)$. 
Here $v_{_{\rm EW}} = 246\,\GeV$ is realized 
when the cutoff scale of NJL dynamics is set as $\Lambda \simeq 66, 223, 480\,\TeV$ 
for $G''/G = 0.3,0.5,0.7$. 
\renewcommand\thefootnote{\alph{footnote}}\footnotetext[1]{
We thank H.~C. Cheng for pointing out the error omitting this obvious condition.
}
\renewcommand\thefootnote{\arabic{footnote}}
%

\section{Summary and Discussion}
\label{summary}

In the spirit of the top quark condensation, 
we proposed a model which has a naturally light composite Higgs boson 
to be identified with the 126 GeV Higgs discovered at the LHC. 
The tHiggs, a bound state of the top quark and its flavor (vector-like) partner, 
emerges as a pseudo Nambu-Goldstone boson (NGB), Top-Mode Pseudo, 
together with the exact NGBs to be absorbed into the $W$ and $Z$ bosons 
as well as another (heavier) Top-Mode Pseudo (CP-odd composite scalar, $A^0_t$).  
Those five composite (exact/pseudo) NGBs are dynamically produced simultaneously 
by a single supercritical four-fermion interaction having 
the $U(3)_{\tilde{\psi}_L} \times U(1)_{\chi_R}$ symmetry which includes the electroweak symmetry, 
where the vacuum is aligned by small explicit breaking term 
so as to break the symmetry down to a subgroup, 
$U(2)_{\tilde{\psi}_L} \times U(1)_{V=\chi_L + \chi_R}$, 
in a way not to retain the electroweak symmetry, 
in sharp contrast to the little Higgs models. 

The $h^0_t$ couplings to the SM particles coincide with those of the SM Higgs boson 
in the limit $\sin \theta = v_{_{\rm EW}}/f \to 0$ with $v_{_{\rm EW}}$ 
being finite (Eqs.(\ref{Model-higgs-coupling}) and (\ref{SMlimit})). 
Even if the tHiggs coupling coincides with that of the SM Higgs, 
the virtue of our model is that the tHiggs $h^0_t$ is a bound state of the top quark and $\chi$-quark, 
and is natural in the sense that its mass is protected by the symmetry, 
in sharp contrast to the SM Higgs. 
 
One notable feature of our model is the prediction of the heavy CP-odd Higgs 
(without additional charged heavy Higgs in contrast to the two-doublet Higgs models). 
The mass of the $A^0_t$ is related to the tHiggs mass (Eq.(\ref{ht-At-mass-relation})) 
at the tree-level of perturbations with respect to the explicit breaking effects,
involving the size of deviation of couplings $(\sin\theta)$ to the electroweak gauge bosons 
from the SM Higgs ones. 
The CP-odd Top-Mode Pseudo $A^0_t$ does not couple to the $W$ and $Z$ bosons 
due to the CP-symmetry and the couplings to other SM particles are generically suppressed 
by $\sin\theta (< 0.34)$. 
Hence the $A^0_t$ is distinguishable from that of the SM-like Higgs boson 
in the high-mass SM Higgs boson search at the LHC.

As noted around Eq.(\ref{quadratic-divergence-mh:meson-loop}) 
the tHiggs would get the significant corrections of higher order 
in $G''/G$ to the mass from 
the top and $t'$-quark as well as the Top-Mode pseudos' loops. 
In particular, the most sizable corrections would come from the top and $t'$-loops 
only at sub-leading order ${\cal O} ((G''/G)^2)$: 
Those one-loop corrections are dominated by the quadratic divergent terms as follows 
(for details of the computations, see Appendix.~\ref{app-toploop-TMP}):
\beq
\left. m^2_{h^0_t} \right|^{t,t'}_{\rm 1-loop}
&=& 
m^2_{h_t^0} 
+ 
\frac{N_c}{(4\pi)^2} y^2 \left( \frac{G''}{G}\right)^2 (2 \cos^2\theta -1) \Lambda^2_\chi
\nonumber \\ 
&=& 
m^2_{h^0_t}
+
\frac{N_c}{8 \pi^2} 
\left( \frac{\sqrt{2} m_t}{v_{_{\rm EW}}} \right)^2 
\left( \frac{2\cos^2\theta-1}{\cos^2\theta} \right)
\Lambda^2_\chi 
\left[ 
1 +{\cal O}\left(\left( \frac{G''}{G}\right)^2 \right)
\right]
\,, \label{quadratic-divergence-mh}
\eeq
where 
the first line is an exact result without further higher order corrections in $G''/G$ 
and
in the second line we used 
$m^2_t = (y^2f^2/2) \sin^2\theta \cos^2\theta (G''/G)^2 [1 + {\cal O}((G''/G)^2)]$ 
which can be derived from Eq.(\ref{Model-tmass}) with Eq.(\ref{def-yukawa})
and $v_{_{\rm EW}} = f \sin\theta$, 
and the chiral symmetry breaking scale $\Lambda_\chi \sim m_{H^0_t}$   
is the cutoff of the nonlinear sigma model constructed from 
Eqs.(\ref{NLsM-Lag-Op2}), (\ref{NLsM-Lag-Op2-mass}), (\ref{NLsM-Lag-Op2-withtopmass}) 
and (\ref{yukawa-no-top}). 
From Eq.(\ref{quadratic-divergence-mh}), 
we see that the perturbative $G''/G$ corrections contribute to the tHiggs mass 
at the order of ${\cal O}((G''/G)^2)$, 
in accord with the Dashen formula Eq.(\ref{Dashen-Formula-G''}).  
Note the amazing cancellation of the quadratic divergent terms among the top and $t'$-quark loops 
when the mixing angle $\theta$ reaches an ideal amount,  
$ \cos\theta = 1/\sqrt{2}$. 
However, to be consistent with the current Higgs coupling measurement at the LHC, 
the $\theta$ is actually strongly constrained to be $\cos\theta > 0.94$ (see Eq.(\ref{bound-theta})), 
which is somewhat far from $\cos\theta = 1/\sqrt{2} \simeq 0.71$. 
Thus the ideal mixing cannot reproduce the reality.

One possibility to make the present model realistic 
would be to pull the $t'$-quark mass down to a low scale in such a way that 
the $t'$-quark can be integrated out. 
In that case, 
we may take the $t'$-quark mass to be the cutoff 
of the nonlinear sigma model, 
$\Lambda_\chi$, say $\Lambda_\chi = m_{t'} \simeq 1.2\,\TeV$  
which is consistent 
with the $S,T$-parameter constraints in Eq.(\ref{ST-t'}) for $\cos\theta = 0.952$. 
Then the one-loop quadratic divergent corrections only come from the tHiggs and $A^0_t$ loops 
as in Eq.(\ref{quadratic-divergence-mh:meson-loop}) 
and top loops involving some effective $h^0_t$-$h^0_t$-$t$-$t$ vertices 
induced from integrating out 
the $t'$-quark (see Appendix.~\ref{app-toploop-TMP}).  
Thus we find the mass shift (with $\Lambda_\chi$ replaced by $m_{t'}$),  
\beq
\left. m^2_{h_t^0} \right|^{t, h^0_t, A^0_t}_{\rm 1-loop}
=
m^2_{h^0_t} \left[ 
1 
 + 
\frac{m^2_{t'}}{(4\pi)^2 v^2_{_{\rm EW}}} \frac{23}{16} 
\right] 
- 
\frac{3}{8\pi^2} \left( \frac{\sqrt{2} m_t}{v_{_{\rm EW}}} \right)^2 
\frac{1 - 6 \cos^2\theta + 6 \cos^4 \theta}{\cos^2\theta} m^2_{t'}
\left[ 
1 +{\cal O}\left(\left( \frac{G''}{G}\right)^2 \right)
\right]
\,. \label{t-prime-integ-mh2}
\eeq
Note the negative correction from the top quark for $\cos\theta = 0.952$,  
in the absence of the $t'$-quark loop contributions. 
Note also that  the result differs from that coming only from the top loop, 
since the $t'$-quark effects are not totally decoupled via equation of motion
as manifested in the induced vertex which never exists in the theory having no $t'$-quark from the onset. 
We thus achieve the desired tHiggs mass around $\simeq 126\,\GeV$ at the one-loop level,  
when the cutoff $\Lambda_\chi = m_{t'}$ is set to $\simeq 1.2\,\TeV$ 
in which case we have the tree-level mass $m_{h^0_t}|_{\rm tree} \simeq 200\,\GeV$. 
 
As seen from the explicit one-loop computation for the $A^0_t$ mass 
given in Appendix.~\ref{app-toploop-TMP}, 
the mixing angle $\cos\theta = 0.952$ is large 
enough to highly suppress the one-loop corrections 
to the $A^0_t$ mass, 
so that we may take 
$m_{A^0_t}|_{\rm tree} \simeq m_{A^0_t}|^{t, h^0_t, A^0_t}_{\rm 1-loop}$ 
from Eq.(\ref{t-prime-integ-mh2}). 
Using the tree-level mass relation among $h^0_t$ and $A^0_t$ 
together with the tree-level tHiggs mass $\simeq 200\,\GeV$, 
we then find 
\beq
\left. m_{A^0_t} \right|_{\rm tree} 
\simeq 
\left. m_{A^0_t} \right|^{t, h^0_t, A^0_t}_{\rm 1-loop} 
\simeq 700\,\GeV 
\,.
\eeq

The scenario in the above would be phenomenologically interesting, 
where the Top-Mode Pseudos ($h^0_t,A^0_t$) 
(and heavy top Higgs $H^t_0$ with the mass $\simeq 2 m_{t'} \simeq 2.4\,\TeV$)   
as well as the $t'$-quark have masses accessible at the LHC. 
Actually, the direct searches for $t'$-quark at the LHC%
~\cite{ATLAS:2013ima,ATLAS-CONF-2013-056,TheATLAScollaboration:2013sha,CMS:2013tda}, 
have placed the limit, $m_{t'} \geq 1 \,\TeV$,  
which is available also to the $t'$-quark in the present model.   
However, 
in addition to usual $t'$-quark searches 
as reported in \cite{ATLAS:2013ima,ATLAS-CONF-2013-056,TheATLAScollaboration:2013sha,CMS:2013tda}, 
a decay channel $t' \to t A^0_t$ would be a characteristic signature 
of the $t'$-quark in the present model.
More on the detailed phenomenological study 
is to be pursued in the future. 

To summarize, the present model predicts the following five masses:    
\beq
m^2_{H^0_t} \text{[Eq.(\ref{mass-Ht0})]}
\quad , \quad 
m^2_{A^0_t} \text{[Eq.(\ref{mass-At0})]}
\quad , \quad
m^2_{h^0_t} \text{[Eqs.(\ref{mass-ht0}) and (\ref{t-prime-integ-mh2})]}
\quad , \quad
m^2_{t'} \text{[Eq.(\ref{Model-t'mass})]}
\quad , \quad
m^2_t \text{[Eq.(\ref{Model-tmass})]}
\,.\label{model-outputs}
\eeq 
These masses are controlled by the five model parameters: 
\beq
G 
\quad , \quad 
G'
\quad , \quad
\Delta_{\chi\chi}
\quad , \quad 
G''
\quad, \quad
\Lambda
\,.
\label{model-parameters}
\eeq
By tuning these model parameters, we can thus realize the mass hierarchies:
\beq
\Lambda ^2
\quad > \quad 
m^2_{H^0_t} \sim m^2_{t'} 
\quad \gg \quad
m^2_{A^0_t} 
\quad > \quad 
m^2_{h^0_t} \sim m^2_t
\,.\label{Model-mass-hierarchy}
\eeq 
The hierarchy $m^2_{H^0_t} \gg m^2_{A^0_t}$ is realized 
by tuning $(0<)G'/G \ll 1$ as in Eq.(\ref{mass-At0}). 
The CP-even Top-Mode Pseudo (tHiggs) mass $m_{h^0_t}$ is smaller than 
the CP-odd Top-Mode Pseudo mass $m_{A^0_t}$ due to the mass relation in Eq.(\ref{mass-ht0}). 
The fermion mass hierarchy $m_{t'} > m_t$ is realized by taking $(0 <) G''/G < 1$ 
(see Eqs.(\ref{Model-t'mass}) and (\ref{Model-tmass})). 

When we set $\Lambda =480 \,\TeV$ and take $G/G_\crit -1 \simeq 10^{-4}$ to get 
$ m_{t'} \simeq 1.2\,\TeV\,, m_{H^0_t} \simeq 2 m_{t'} \simeq 2.4 \, {\rm TeV}$.
Taking $v_{_{\rm EW}} = 246 \,\GeV$, 
which determines $m_{t\chi}$ through Eq.(\ref{def-vEW}), 
and setting $G'/G \simeq 10^{-5}$ and $G''/G \simeq 0.7$, 
we have $m_t \simeq 173 \,\GeV$ and 
$m_{h^0_t}|^{t, h^0_t, A^0_t}_{\rm 1-loop} \simeq 126 \,\GeV$ 
where $m_{h^0_t}|_{\rm tree} \simeq 200 \,\GeV$.
Thus the phenomenologically favored situation as above can be realized 
from the original four-fermion dynamics.

\section*{Acknowledments}
We would like to thank Hiroshi Ohki for useful discussions. 
This work was supported by the JSPS Grant-in-Aid for Scientific Research (S) \#22224003 and (C) \#23540300 (K.Y.).

\appendix
\section{Alternative derivation of Top-Mode Pseudo mass formulas} 
\label{app-BHL}

In this appendix we derive the mass formulas for the Top-Mode Pseudos 
in Eqs.(\ref{mass-At0}) and (\ref{mass-ht0}) 
based on the Bardeen-Hill-Lindner approach~\cite{Bardeen:1989ds}. 

Introducing the auxiliary fields 
$\phi_{t\chi}\sim \bar{\chi}_R t_L$,  $\phi_{b\chi}\sim \bar{\chi}_R b_L$ 
and $\phi_{\chi\chi}\sim \bar{\chi}_R \chi_L$, 
we rewrite the Lagrangian Eq.(\ref{with-explicit-breaking-00})
into a linear sigma model-like form including the $1/N_c$-leading corrections renormalized 
at the scale $\mu (< \Lambda)$, 
\beq 
{\cal L}_{\rm BHL}
&=& 
{\cal L}_{\rm kin.}
-
\frac{1}{\sqrt{Z}}
\left[ 
\bpm \bar{t}_L & \bar{b}_L\epm \bpm \phi_{t\chi} \\ \phi_{b\chi}\epm\chi_R 
+ \bar{\chi}_L \phi_{\chi\chi}\chi_R
+ \text{h.c.}
\right]
+ {\cal L}_{\text{L$\sigma$M}}
\,,\label{model-BHL-full}
\eeq
where 
\beq
{\cal L}_{\text{L$\sigma$M}}
&=&
\left| D_\mu \bpm \phi_{t\chi} \\[0.5ex] \phi_{b\chi} \epm \right|^2
+
\left| \partial_\mu \phi_{\chi\chi}\right|^2
-V(\phi)
\,,\label{model-LsM}
\\
D_\mu \bpm \phi_{t\chi} \\[0.5ex] \phi_{b\chi} \epm
&=&
\left( \partial_\mu -i g W^a_\mu \frac{\sigma^a}{2} + i g' \frac{1}{2} B_\mu \right) 
\bpm \phi_{t\chi} \\[0.5ex] \phi_{b\chi} \epm
\,,\nonumber
\eeq
and 
\beq
V(\phi)
&=&
M^2\left[\phi^\dagger_{t\chi}\phi_{t\chi} + \phi^\dagger_{b\chi}\phi_{b\chi} + \phi^\dagger_{\chi\chi}\phi_{\chi\chi}  \right]
+
\lambda \left[ 
\phi^\dagger_{t\chi}\phi_{t\chi} + \phi^\dagger_{b\chi}\phi_{b\chi} + \phi^\dagger_{\chi\chi}\phi_{\chi\chi}
\right]^2
\nonumber\\
&&
+\Delta M^2 \phi^\dagger_{\chi\chi} \phi_{\chi \chi}
- C_{\chi\chi} \left[ \phi^\dagger_{\chi \chi} + \phi_{\chi\chi}\right]
\,,\label{model-LsM-V}
\eeq
\beq
Z &=& \frac{1}{\lambda}= \frac{N_c}{16 \pi^2} \ln \frac{\Lambda^2}{\mu^2}
\quad , \quad
M^2 = \frac{1}{Z} \left(\frac{1}{G} - \frac{N_c}{8\pi^2} \Lambda^2\right)
\,,\,
\,\nonumber\\[1ex]
\Delta M^2 &=& 
\frac{1}{Z} \left(\frac{1}{G-G'} - \frac{1}{G} \right)
\quad ,\quad
C_{\chi\chi} =
\frac{1}{\sqrt{Z}} \frac{\Delta_{\chi\chi}}{G-G'}
\,.\nonumber
\eeq
We define the vacuum expectation values corresponding to Eq.(\ref{condensation-psiBasis}),  
\beq
\vev{\phi_{t\chi}} = \frac{f \sin\theta}{\sqrt{2}} \equiv \frac{v_{t\chi}}{\sqrt{2}}
\quad , \quad
\vev{\phi_{b\chi}} = 0
\quad , \quad
\vev{\phi_{\chi\chi}} = \frac{f\cos\theta}{\sqrt{2}} \equiv \frac{v_{\chi\chi}}{\sqrt{2}}
\,,\label{vev-TSS-LsM}
\eeq
and the dynamical masses, 
\beq
m_{AB} = \frac{1}{\sqrt{Z}} \vev{\phi_{AB}}
\quad \text{for} \quad 
A, B = t,b,\chi 
\,.\label{dynamical-mass-BHL}
\eeq 
The stationary conditions, corresponding to the gap equations 
Eqs.(\ref{gapeq-tchi}) and (\ref{gapeq-chichi}), are obtained  from the potential $V(\phi)$ to be 
\beq
&&
\frac{\partial V}{\partial v_{t\chi}} = 0
\quad \Leftrightarrow \quad
m_{t\chi} 
=
m_{t\chi} \frac{N_c G }{8\pi^2} 
\left[
\Lambda^2 -
\left( m^2_{t\chi} + m^2_{\chi \chi} \right) \ln \frac{\Lambda^2}{m^2_{t\chi} + m^2_{\chi\chi}}
\right]
\,,\label{gapeq-tchi-BHL}
\\[1ex]
&&
\frac{\partial V}{\partial v_{\chi\chi}} = 0
\quad \Leftrightarrow \quad
m_{\chi\chi} 
=
\Delta_{\chi\chi}
+
m_{\chi\chi} \frac{N_c (G-G') }{8\pi^2} 
\left[
\Lambda^2 -
\left( m^2_{t\chi} + m^2_{\chi \chi} \right) \ln \frac{\Lambda^2}{m^2_{t\chi} + m^2_{\chi\chi}}
\right]
\,.\label{gapeq-chichi-BHL}
\eeq

We next parametrize the neutral scalar fields $\phi_{t\chi}$ and $\phi_{\chi\chi}$ as
\beq
\phi_{t\chi} = \frac{v_{t\chi} + \text{Re\,}\phi_{t\chi} + i \text{Im\,}\phi_{t\chi}}{\sqrt{2}}
\quad , \quad
\phi_{\chi\chi} = \frac{v_{\chi\chi} + \text{Re\,}\phi_{\chi\chi} + i \text{Im\,}\phi_{\chi\chi}}{\sqrt{2}}
\,.
\eeq
Taking into account the stationary conditions Eqs.(\ref{gapeq-tchi-BHL}) and (\ref{gapeq-chichi-BHL}), 
we find the mass terms of 
$(\text{Re} \phi_{t\chi},\text{Re} \phi_{\chi\chi},\text{Im} \phi_{\chi\chi})$ 
in the effective potential Eq.(\ref{model-LsM-V}), 
\beq
-\frac{1}{2} m^2_{A^0_t} (\text{Im\,}\phi_{\chi\chi})^2
-\frac{1}{2} 
\bpm \text{Re\,}\phi_{t\chi} & \text{Re\,}\phi_{\chi\chi} \epm
\bpm
4m^2_{t\chi} & 4m_{t\chi} m_{\chi\chi} \\
4m_{t\chi} m_{\chi\chi} & 4 m^2_{\chi\chi}+ m^2_{A^0_t}
\epm
\bpm \text{Re\,}\phi_{t\chi} \\ \text{Re\,}\phi_{\chi\chi} \epm
\,, \label{mass-matrix-LsM}
\eeq
where $\text{Im} \phi_{\chi\chi} \equiv A^0_t$ 
and the $A^0_t$ mass $m_{A^0_t}$ is given by expanding terms in powers of $G'/G \ll 1$ as 
\beq 
m^2_{A^0_t} &=&
\Delta M^2
\nonumber\\
&=&
\left( \frac{1}{G-G'} - \frac{1}{G}\right) 
\times \frac{16\pi^2}{N_c\ln(\Lambda^2/ ( m^2_{t\chi} + m^2_{\chi\chi}))}
\nonumber\\
&\simeq&
\frac{16\pi^2}{GN_c\ln(\Lambda^2/ ( m^2_{t\chi} + m^2_{\chi\chi}))}
\left(\frac{G'}{G}\right)
\left[1 + {\cal O}\left( \left( \frac{G'}{G}\right)^2\right) \right]
\nonumber \\ 
&\simeq& 
\frac{2(m^2_{t\chi} + m^2_{\chi\chi})}{Gf^2} 
\left(\frac{G'}{G}\right) 
\left[1 + {\cal O}\left( \left( \frac{G'}{G}\right)^2\right) \right]
\,.\label{LsM-mA}
\eeq 
where the renormalization scale $\mu$ has been set to $(m^2_{t\chi} + m^2_{\chi\chi})^{1/2}$ 
and use has been made of the Pagels-Stokar formula for the decay constant $f$ given 
in Eq.(\ref{model-PSformula}). 
For $G'/G\ll 1$ 
we may take $m_{A^0_t} \ll m_{\tilde{\chi}\chi} = (m^2_{t\chi} + m^2_{\chi\chi})^{1/2}$ 
so that the scalar mass in the last term of Eq.(\ref{mass-matrix-LsM}) 
can be diagonalized 
up to terms of ${\cal O}(m^4_{A^0_t}/m^2_{\tilde{\chi}\chi})$ as
\beq
\bpm
4m^2_{t\chi} & 4m_{t\chi} m_{\chi\chi} \\[1ex]
4m_{t\chi} m_{\chi\chi} & 4 m^2_{\chi\chi}+ m^2_{A^0_t}
\epm
\simeq
\bpm \cos \theta & \sin\theta \\[1ex] -\sin\theta & \cos\theta\epm
\bpm
m^2_{A^0_t} \sin^2\theta & 0 \\[1ex]
0 & 4 (m^2_{t\chi} + m^2_{\chi\chi})+m^2_{A^0_t} \cos^2\theta
\epm
\bpm \cos \theta & -\sin\theta \\[1ex] \sin\theta & \cos\theta\epm
\,,\nonumber
\eeq
where $\tan \theta \equiv m_{t\chi}/m_{\chi\chi}$ 
and the corresponding mass eigenstates $h^0_t$ and $H^0_t$ with $m_{H^0_t} > m_{h^0_t}$ are given as 
\beq
\bpm h^0_t \\[1ex] H^0_t \epm
=
\bpm \cos \theta & -\sin\theta \\[1ex] \sin\theta & \cos\theta\epm
\bpm \text{Re\,}\phi_{t\chi} \\[1ex] \text{Re\,}\phi_{\chi\chi} \epm
\,.\nonumber
\eeq
Thus we find the masses of the two CP-even neutral scalar mesons 
up to terms of ${\cal O}(m^4_{A^0_t}/m^2_{\tilde{\chi}\chi})$:
\beq
m^2_{h^0_t} \simeq m^2_{A^0_t} \sin^2\theta 
\quad , \quad
m^2_{H^0_t} \simeq 4 (m^2_{t\chi} + m^2_{\chi\chi})
\,.
\label{LsM-mh}
\eeq
Eqs.(\ref{LsM-mA}) and (\ref{LsM-mh}) exactly reproduce 
the Top-Mode Pseudo mass formulas in Eqs.(\ref{mass-At0}) and (\ref{mass-ht0}) 
obtained from the nonlinear Lagrangian with the heavy top-Higgs integrated out.

\section{$t,t'$-loop corrections to the Top-Mode Pseudo masses} 
\label{app-toploop-TMP}

In this appendix, 
we shall compute the quadratic divergent corrections to the Top Mode Pseudos $(h^0_t, A^0_t)$ 
arising from the top and $t'$-loops as the $1/N_c$-leading contribution 
in Eqs.(\ref{quadratic-divergence-mh}) and (\ref{t-prime-integ-mh2}).  

\begin{figure}[htbp]
\begin{center}
\includegraphics[scale=0.6]{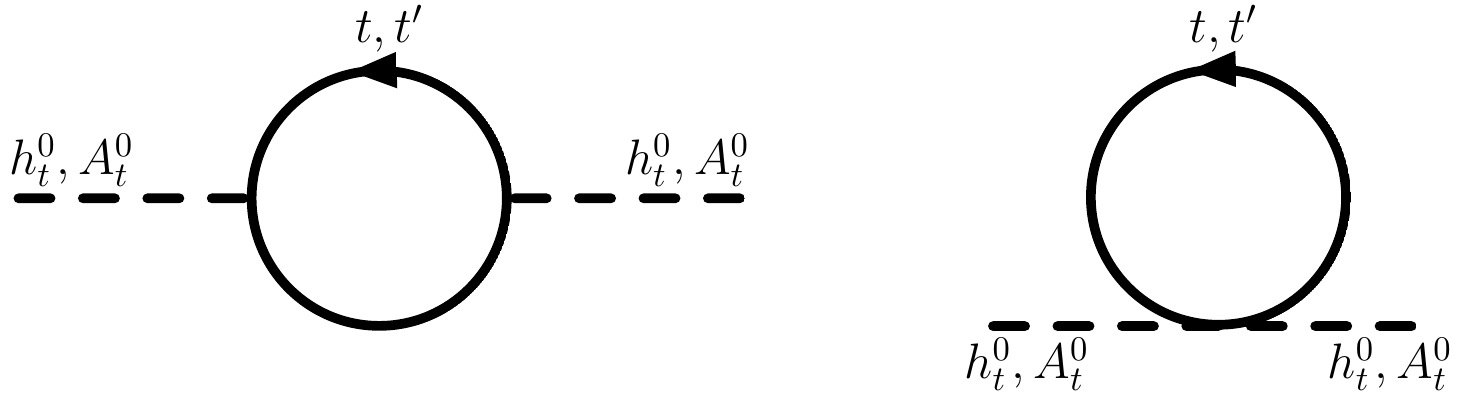}
\caption{
The one-loop diagrams contributing to $h^0_t,A^0_t$ masses 
as the quadratic divergent corrections up to ${\cal O}((G''/G)^2)$. 
\label{one-loop-graphs}
}
\end{center} 
\end{figure}

We start with 
the Yukawa sector in Eq.(\ref{NLsM-Lag-Op2-withtopmass}), 
\beq
{\cal L}^{t,t'}_{\rm yuk.}
=
-
\frac{f}{\sqrt{2}}
\left[ 
y \bar{\psi}_L (R^T U) \psi_R
+
y_{\chi t}\bar{\psi}_L (\chi_1 R^T U \chi_3) \psi_R
+ \text{h.c.}
\right]
\,.
\eeq 
From this Lagrangian, 
we find the couplings relevant to the one-loop corrections 
in the basis of the mass-eigenstates $(t, t')_m$: 
\beq
\left.
{\cal L}^{t,t'}_{\rm yuk.} 
\right|_{h^0_t}
&=&
-y_{htt} h^0_t \bar{t}t 
-y_{ht't'}  h^0_t \bar{t'}t'
-g_{ht_Lt'_R} h^0_t \left( \bar{t}_L t'_R +\text{h.c.} \right)
-g_{ht_Rt'_L} h^0_t \left( \bar{t}_R t'_L +\text{h.c.} \right)
\nonumber\\
&&
-g_{hhtt}h^0_th^0_t\bar{t}t
-g_{hht't'}h^0_th^0_t\bar{t}'t'
+ \cdots
\,,\label{yukawa-coupling}
\eeq
and 
\beq
\left.
{\cal L}^{t,t'}_{\rm yuk.} 
\right|_{A^0_t}
&=&
-y_{Att} A^0_t \bar{t}\gamma_5 t 
-y_{At't'}  A^0_t \bar{t'} \gamma_5 t'
-g_{At_Lt'_R} A^0_t  \bar{t} \gamma_5 t'
-g_{At_Rt'_L} A^0_t  \bar{t} \gamma_5 t' 
\nonumber\\
&&
-g_{AAtt}A^0_t A^0_t\bar{t}\gamma_5 t
-g_{AAt't'}A^0_t A^0_t\bar{t}'\gamma_5 t'
+ \cdots
\,,\label{yukawa-coupling}
\eeq
where 
\beq
y_{htt} 
&=& 
\frac{y}{\sqrt{2}} \left[ 
(c^t_L \cos \theta  + s^t_L \sin \theta ) s^t_R - s^t_Lc^t_R \sin\theta \left( \frac{G''}{G} \right)
\right]
\,,\label{gTM-htt}\\
y_{ht't'} 
&=& 
\frac{y}{\sqrt{2}}  \left[ 
( s^t_L \cos \theta - c^t_L \sin \theta) c^t_R - c^t_L s^t_R \sin\theta \left( \frac{G''}{G} \right)
\right]
\,,\label{gTM-ht't'}\\
g_{ht_Lt'_R}
&=& 
\frac{y}{\sqrt{2}}\left[
(c^t_L \cos\theta + s^t_L \sin\theta) c^t_R + s^t_L s^t_R \sin\theta \left( \frac{G''}{G}\right)
\right]
\,,\label{gTM-htLt'R}\\
g_{ht'_Lt_R}
&=& 
\frac{y}{\sqrt{2}} \left[ 
(s^t_L \cos\theta - c^t_L \sin\theta) s^t_R + c^t_L c^t_R \sin\theta \left( \frac{G''}{G} \right)
\right]
\,,\label{gTM-ht'LtR}\\
g_{hhtt}
&=& 
\frac{y}{2\sqrt{2} f}\left[ 
(c^t_L \sin\theta - s^t_L \cos\theta) s^t_R 
+ s^t_L c^t_R \cos\theta \left( \frac{G''}{G}\right) 
\right]
\,,\label{gTM-hhtt}\\
g_{hht't'}
&=& 
\frac{y}{2\sqrt{2} f}\left[ 
(s^t_L \sin\theta + c^t_L \cos\theta) c^t_R 
+ c^t_L s^t_R \cos\theta \left( \frac{G''}{G}\right) 
\right]
\,,\label{gTM-hht't'}
\eeq
and 
\beq
y_{Att} 
&=& 
\frac{iy}{\sqrt{2}} \left[ 
-s^t_L s^t_R + s^t_Lc^t_R \left( \frac{G''}{G} \right)
\right]
\,,\label{gTM-Att}\\
y_{At't'} 
&=& 
\frac{iy}{\sqrt{2}}  \left[ 
c^t_L c^t_R + c^t_L s^t_R \left( \frac{G''}{G} \right)
\right]
\,,\label{gTM-At't'}\\
g_{At_Lt'_R}
&=& 
\frac{iy}{\sqrt{2}}\left[
-s^t_L c^t_R - s^t_L s^t_R \left( \frac{G''}{G}\right)
\right]
\,,\label{gTM-AtLt'R}\\
g_{At'_Lt_R}
&=& 
\frac{iy}{\sqrt{2}} \left[ 
c^t_L s^t_R - c^t_L c^t_R \left( \frac{G''}{G} \right)
\right]
\,,\label{gTM-At'LtR}\\
g_{AAtt}
&=& 
g_{hhtt}
+
\frac{3y}{4\sqrt{2} f}\sin\theta\cos\theta
\left[
-(c^t_L \cos\theta +s^t_L \sin\theta)s^t_R + s^t_Lc^t_R \sin\theta \left( \frac{G''}{G}\right)
\right]
\,,\label{gTM-AAtt}\\
g_{AAt't'}
&=& 
g_{hht't'}
+
\frac{3y}{4\sqrt{2} f}\sin\theta\cos\theta
\left[
-(s^t_L \cos\theta -c^t_L \sin\theta)c^t_R + c^t_Ls^t_R \sin\theta \left( \frac{G''}{G}\right)
\right]
\,. 
\label{gTM-AAt't'}
\eeq
The one-loop corrections arise from Feynman graphs involving the top and $t'$-loops 
as depicted in Fig.~\ref{one-loop-graphs}. 
The quadratic divergent corrections to the Top-Mode Pseudo masses are thus calculated to be 
\beq
\delta m^2_{h^0_t,A^0_t}
= 
\frac{N_c }{8\pi^2} \Lambda^2_\chi 
\cdot {\cal C}_{h,A} 
\,,\label{TMP-mass-correction}
\eeq
where the chiral symmetry breaking scale $\Lambda_\chi$   
is the cutoff of the nonlinear sigma model constructed from 
Eqs.(\ref{NLsM-Lag-Op2}), (\ref{NLsM-Lag-Op2-mass}), (\ref{NLsM-Lag-Op2-withtopmass}) 
and (\ref{yukawa-no-top}) 
and
\beq
{\cal C}_h
&=&
-y^2_{htt}-y^2_{ht't'}-g^2_{ht_Lt'_R}-g^2_{ht'_Lt_R}
+ g^2_{hhtt} + g^2_{hht't'}
\nonumber\\
&=&
\frac{y^2}{2}
\left( \frac{G''}{G}\right)^2 
(2\cos^2\theta-1)
\,,\\
{\cal C}_A
&=&
-y^2_{Att}-y^2_{At't'}-g^2_{At_Lt'_R}-g^2_{At'_Lt_R}
+ g_{hhtt} g_{AAtt} + g_{hht't'} g_{AAt't'}
\nonumber\\
&=&
\frac{y^2}{2}
\left( \frac{G''}{G}\right)^2 
\frac{1}{2}(1-\cos^2\theta)(3\cos^2\theta-2)
\,.
\eeq
Here we used the orthogonality relations 
among the mixing angles $s^t_{L,R}$ ($c^t_{L,R}$) 
which follows from the diagonalization of the fermion mass matrix in Eq.(\ref{mass-ttprime}) 
with the rotation matrices in Eq.(\ref{rotate-ttprime}):   
\beq
s^t_L s^t_R \cos\theta \left( \frac{G''}{G} \right) 
= \left(c^t_L \sin\theta - s^t_L \cos\theta \right) c^t_R 
\,, \qquad 
c^t_L c^t_R \cos\theta \left( \frac{G''}{G} \right) 
= \left( s^t_L \sin\theta + c^t_L \cos\theta \right) s^t_R 
\,. 
\eeq
Thus we have Eq.(\ref{quadratic-divergence-mh}) 
and the associated formula for $A^0_t$:
\beq
\left. \delta m^2_{h^0_t} \right|^{t,t'}
&=& 
\frac{N_c}{8\pi^2} 
\frac{y^2}{2}
\left( \frac{G''}{G}\right)^2 
(2\cos^2\theta-1)
\Lambda^2_\chi
\nonumber\\
&=&
\frac{N_c}{8 \pi^2} 
\left( \frac{\sqrt{2} m_t}{v_{_{\rm EW}}} \right)^2 
\left( \frac{2\cos^2\theta-1}{\cos^2\theta}\right)\Lambda^2_\chi 
\left[ 
1 +{\cal O}\left(\left( \frac{G''}{G}\right)^2 \right)
\right]
\nonumber\\
&=&
\text{Last term in Eq.(\ref{quadratic-divergence-mh})}
\label{h-tt'-loop}
\,, \\ 
\left. \delta m^2_{A^0_t} \right|^{t,t'}
&=& 
\frac{N_c}{8\pi^2} 
\frac{y^2}{2}
\left( \frac{G''}{G}\right)^2 
\frac{1}{2}(1-\cos^2\theta)(3\cos^2\theta-2)
\Lambda^2_\chi
\nonumber\\
&=&
\frac{N_c}{8 \pi^2} 
\left( \frac{\sqrt{2} m_t}{v_{_{\rm EW}}} \right)^2 
\frac{(1-\cos^2\theta)(3\cos^2\theta-2)}{2\cos^2\theta}\Lambda^2_\chi
\left[ 
1 +{\cal O}\left(\left( \frac{G''}{G}\right)^2 \right)
\right]
\,,\label{A-tt'-loop}
\eeq
where 
the first lines in Eqs.(\ref{h-tt'-loop}) and (\ref{A-tt'-loop}) 
are exact results without further higher order corrections in $G''/G$ 
and
in the second lines in Eqs.(\ref{h-tt'-loop}) and (\ref{A-tt'-loop}) 
we used　
$m^2_t = (y^2f^2/2) \sin^2\theta \cos^2\theta (G''/G)^2 [1 + {\cal O}((G''/G)^2)]$ 
which can be derived from Eq.(\ref{Model-tmass}) with Eq.(\ref{def-yukawa})
and $v_{_{\rm EW}} = f \sin\theta$. 
As noted around Eq.(\ref{quadratic-divergence-mh}), 
from Eq.(\ref{h-tt'-loop}), 
we see that the perturbative $G''/G$ corrections contribute to the tHiggs mass 
at the order of ${\cal O}((G''/G)^2)$, 
no correction of ${\cal O}(G''/G)$, 
in accord with the Dashen formula Eq.(\ref{Dashen-Formula-G''}), 
as well as the $A^0_t$ mass in Eq.(\ref{A-tt'-loop}).

In the limit where $m_t \ll m_{t'} \sim \Lambda_\chi$,   
the $t'$-quark may be integrated out to induce  
the effective $h^0_t$-$h^0_t$-$t$-$t$ and $A^0_t$-$A^0_t$-$t$-$t$-couplings, 
\beq
g'_{hhtt}
= \frac{y}{2\sqrt{2}f} \frac{g_{ht_Lt'_R} g_{ht'_L t_R}}{g_{hht't'}}
\quad, \quad
g'_{AAtt}
= \frac{-y}{2\sqrt{2}f} \frac{g_{At_Lt'_R} g_{At'_L t_R}}{g_{hht't'}}
\,. 
\eeq
In that case, ${\cal C}_{h,A}$ in Eq.(\ref{TMP-mass-correction}) become, 
up to order of $(G''/G)^2$,
\beq
{\cal C}_h
&=&
-y^2_{htt}+ g^2_{hhtt} + g_{hhtt}g'_{hhtt}
\nonumber\\
&=&
\frac{y^2}{2}
\left( \frac{G''}{G}\right)^2 
(-1+6\cos^2\theta - 6\cos^4\theta)
\left[ 1 + {\cal O}\left( \left( \frac{G''}{G}\right)^2\right)\right]
\,,\\
{\cal C}_A
&=&
-y^2_{Att}+ g^2_{AAtt} + g_{hhtt}g'_{AAtt}
\nonumber\\
&=&
\frac{y^2}{2}
\left( \frac{G''}{G}\right)^2 
\frac{1}{2}(1-\cos^2\theta)(-2+7\cos^2\theta -6\cos^4\theta)
\left[ 1 + {\cal O}\left( \left( \frac{G''}{G}\right)^2\right)\right]
\,. 
\eeq
Thus we find Eq.(\ref{t-prime-integ-mh2}) and 
the associated result for $A^0_t$:
\beq
\left. \delta m^2_{h^0_t} \right|^t
&=& 
\frac{N_c}{8\pi^2} 
\frac{y^2}{2}
\left( \frac{G''}{G}\right)^2 
(-1+6\cos^2\theta - 6\cos^4\theta)
m^2_{t'}
\left[ 1 + {\cal O}\left( \left( \frac{G''}{G}\right)^2\right)\right]
\nonumber\\ 
&=&
- 
\frac{3}{8\pi^2} \left( \frac{\sqrt{2} m_t}{v_{_{\rm EW}}} \right)^2 
\frac{1 - 6 \cos^2\theta + 6 \cos^4 \theta}{\cos^2\theta} m^2_{t'}
\left[ 1 + {\cal O}\left( \left( \frac{G''}{G}\right)^2\right)\right]
\nonumber\\
&=&
\text{Last term in Eq.(\ref{t-prime-integ-mh2})}
\,,\\
\left. \delta m^2_{A^0_t} \right|^t
&=& 
\frac{N_c}{8\pi^2} 
\frac{y^2}{2}
\left( \frac{G''}{G}\right)^2 
\frac{1}{2}(1-\cos^2\theta)(-2+7\cos^2\theta -6\cos^4\theta) m^2_{t'}
\left[ 1 + {\cal O}\left( \left( \frac{G''}{G}\right)^2\right)\right]
\nonumber\\ 
&=&
- 
\frac{3}{8\pi^2} \left( \frac{\sqrt{2} m_t}{v_{_{\rm EW}}} \right)^2 
\frac{(1-\cos^2\theta)(2-7\cos^2\theta +6\cos^4\theta)}{2\cos^2\theta} m^2_{t'}
\left[ 1 + {\cal O}\left( \left( \frac{G''}{G}\right)^2\right)\right]
\,.  
\eeq

\bibliography{Higgs-as-TP}
\end{document}